\documentclass[12pt]{article}
\usepackage[latin1]{inputenc}
\usepackage{graphicx}
\usepackage{float}
\usepackage{amsfonts}
\begin{document}
\title{Confinement of quarks and valence gluons in $SU(N)$ Yang-Mills-Higgs models}
\author{~L.~E.~Oxman\\ \\ Department of Applied Mathematics and Theoretical Physics,\\
University of Cambridge, Cambridge, CB3 OWA, UK
\\ \\ Instituto de F\'{\i}sica, Universidade Federal Fluminense,\\
Campus da Praia Vermelha, Niter\'oi, 24210-340, RJ, Brazil.
}
\date{\today}
\maketitle

\begin{abstract}

In this work, we analyze a class of Yang-Mills models containing adjoint Higgs fields, with $SU(N)$ symmetry spontaneously broken down to $Z(N)$, showing they  contain center vortices, Y-junctions formed by them, and 
junctions where different center vortices are smoothly interpolated by monopole-like configurations.   
 In the context of dual superconductors, these objects represent different states of the gluon field. 
Center vortices confine quarks to form normal hadron states. The interpolating monopole, which in our model cannot exist as an isolated configuration, is identified with a confined valence gluon. 
A junction containing a monopole can bind quarks in a color nonsinglet state to form an overall neutral object, identified with a hybrid hadron. These states, formed by quarks bound to a valence gluon, are allowed by QCD, and current experimental collaborations are aimed  at identifying them. Finally, considering the general version of the model, based on a compact simple gauge group $G$, the picture is completed with a heuristic discussion about why it would be natural using as $G$ the dual of  the chromoelectric gauge group $G_{\rm e}$, and external pointlike monopoles to represent the mesonic and baryonic Wilson loops.

\end{abstract}

{\bf Keywords}: \\

{\bf Pacs}: 11.15.-q, 11.15.Tk

\section{Introduction}

Nowadays, the implementation of sophisticated lattice numerical techniques together with ever-expanding computing capacities is permitting
to unveil the complex dynamics of the strong interactions. 

Using the $\pi$, $K$ and $\Xi$ masses as input parameters,  the obtained light hadron spectrum is in very good agreement with the experimental one \cite{durr}. The properties of the potential between heavy $q\bar{q}$ probes in pure $SU(N)$ Yang-Mills (YM)  theory were obtained for different representations of the external quarks. Among them, linearity and $N$-ality at asymptotic distances, Casimir scaling at intermediate ones, and string-like behavior, have been observed \cite{greensite,book-G}. Recent lattice calculations permitted to elucidate the type of interaction between three heavy quarks in pure $SU(3)$ YM theory. The controversy about whether the $3q$ potential is $\Delta$ or Y-shaped at asymptotic distances was resolved in favor of the latter \cite{suganuma1}-\cite{cornwall}. 

Regarding the properties of the gluon field, there is another  central question in hadron spectroscopy, which is still open, concerning the existence of a color nonsinglet $q\bar{q}' $ pair neutralized in color by an excited gluon field,  a state known as a hybrid meson.  These $qg\bar{q}'$ states, where the quarks are bound to a valence gluon, as well as similar hybrid baryonic states, are allowed by QCD and have been searched for many years now. Among the new experiments, the glueX collaboration at the Jefferson Lab aims at mapping the spectrum of gluonic excitations using polarized photons. For a review of  theoretical predictions for masses and decay modes, recent experimental evidence, and some lattice results, see \cite{hybrids}-\cite{DuE} and references therein.

On the theoretical side, motivated by the pioneering work of refs. \cite{N}-\cite{3}, 
Abelian Higgs models to describe the behavior of the $q\bar{q}$ potential \cite{Bali} and  the interaction among three quarks in normal hadrons, have been proposed in  \cite{Baker,Baker1} and  \cite{MSu}-\cite{C1}, respectively. Many efforts have also been devoted to understand dualities in supersymmetric theories containing non Abelian monopoles \cite{GNO}-\cite{Shifman} and non Abelian vortices \cite{David,it}, for a review, see refs. \cite{notes,vortices-d}. In refs. \cite{abek}-\cite{Dorigoni}, a non Abelian complex formed by an open vortex ending at a monopole/antimon\-o\-pole pair, whose positions are kept fixed, has been obtained in supersymmetric models with hierarchical gauge symmetry breaking. In ref. \cite{Cesar}, a $Z(N)$ $(2+1)d$ model was constructed where not only quark/antiquark pairs are confined by domain walls, which in planar systems are stringlike, but also contains baglike baryonic states. Planar models that admit stable static domain wall junctions have been proposed in ref. \cite{GT}.

In pure $SU(N)$ YM theory, the $N$-ality dependence of the $q\bar{q}$ potential points to center symmetry as a natural ingredient in the possible dual effective models. In particular, heavy quarks in the adjoint representation should not be confined \cite{greensite,book-G}, a property that would be naturally implemented by the triviality of vortices with center charge $N$. In ref. \cite{Konishi-Spanu}, the general discussion about the magnetic weights of non Abelian monopoles \cite{GNO} was adapted to classify center vortices in theories where the gauge group $G$ is spontaneously broken down to its center (for a review, see \cite{notes}). An $SU(N)$ model contaning stable center vortices, with at least $N$ adjoint Higgs fields, has been previously proposed in refs. \cite{deVega}-\cite{HV}. 

In spite of all the above mentioned theoretical activity, a $(3+1)d$ non Abelian model containing the different 
gluon field states needed to form normal and hybrid hadrons, is still lacking.
In this article, we shall propose a first step towards this objective, concentrating on the underlying topological features. Although we will be mainly interested in $SU(N)$ Yang-Mills-Higgs (YMH) dual models, most of our discussion will also apply to a general compact simple $G$ with nontrivial center.

In our model, the normal gluon field will correspond to center vortices, or Y-junctions exclusively formed by them, while the unconventional field, characteristic of either mesonic or baryonic hybrids, to  different center vortices smoothly interpolated by monopole-like configurations, which represent the confined valence gluons. 
In addition, the quark probes will be represented by external pointlike monopole sources.  

Our work is motivated by a different but related aspect of gauge theories, namely, the search for the relevant configurations to compute averages in pure YM. 
In the lattice, some scenarios involving ensembles of magnetic defects exist, having their counterpart in the continuum. Monopoles have been introduced by means of a gauge fixing where a field transforming in the adjoint is diagonalized \cite{ap}. They have also been discussed in a gauge invariant way, in terms of a field decomposition that permits a simple description of monopoles as defects of a local direction $n_1$ in color space \cite{Manton}-\cite{Shaba}.
Thin center vortices have been parametrized in \cite{engelhardt1,reinhardt}. In ref. \cite{lucho},  in the case of $SU(2)$, we have written these singular objects as defects in a pair of local directions $n_2, n_3$, which together with  $n_1$ form a local color frame for the Lie algebra. We also showed this type of description extends to correlated monopoles and center vortices, configurations that in the lattice are known as chains. 

Now, the general discussion about the group weights and the topology for $Z(N)$ vortices in dual superconductor YMH models \cite{Konishi-Spanu} is similar to that for magnetic $Z(N)$ thin center vortices in pure YM \cite{engelhardt1}-\cite{reinhardt}. Therefore, a part of the analysis for YMH hybrid states should be similar to that for YM correlated vortices and monopoles.  As we will see, in the YMH model defined by a gauge group $G$, the defects of a local frame for the Lie algebra $\mathfrak{g}$  will provide the guiding centers for the smooth hybrid configurations. 

For this aim, the Higgs potential will be chosen such that the vacua are labeled by $Ad(G)$, the adjoint representation of $G$. As a concrete example, we shall consider as many adjoint Higgs fields as the dimension of 
$\mathfrak{g}$, propose a flavor symmetry with global group $Ad(G)$, and construct a potential whose terms only involve the product 
in the algebra. However, the general discussion could also be applied to other potentials with the same symmetry breaking pattern, 
and diagonal and off-diagonal sectors, setting a scale for the monopoles and center vortices.
This type of model has also points of contact with the $SU(N)$ YMH model in \cite{Fidel2}, describing center vortices by means 
of adjoint Higgs fields along directions orthogonal with respect to a diagonal sector, although in that case the diagonal sector has been frozen in a way it does not contribute to the energy. The flavor symmetry also plays an important role in models containing non Abelian vortices \cite{David}, constructed with as many fundamental Higgs fields as the dimension of the fundamental representation of $G$. 

Regarding the two different aspects related with center vortices and mon\-o\-poles, namely, their study in either the context of
dual superconductor YMH models with gauge group $G$, or of pure YM theory with chromoelectric gauge group $G_{\rm e}$, 
the following remarks are in order. First, it is important to underline that, in the former case, a single configuration has a direct physical meaning, representing the chromoelectric confining string or junction acting between heavy quark probes. On the other hand, in 
the latter, a single configuration has no direct physical meaning, but  represents possible virtual processes in $4D$ Euclidean spacetime, to be summed in physical objects such as the partition function. Second, while in YMH models the topological objects are smooth and essentially classical, in pure YM theory they are initially singular, but lattice results lead to consider the possibility they could become thick and relevant due to quantum effects \cite{greensite,book-G}. 
The hope is that in pure YM the effective action of magnetic topological configurations become small or negative \cite{diakonov-center}. In that case, they would proliferate and lead to a dual superconductor picture where similar topological objects,  in a Higgs phase, represent the different states of the confining gluon field. For further considerations about this important point, see \S \ref{appa} and refs. \cite{Lucho1}-\cite{Lucho3}.

The article is organized as follows. In \S \ref{YMHm}, we present our model and analyze the possible vacua. In
\S \ref{topo-cent} and \S \ref{chains-comp}, we discuss the topology and charges of center vortices and junctions. 
Some basic properties of the Cartan decomposition of $SU(N)$ are reviewed in \S \ref{rev-sun}. In \S \ref{cva}, we show how to close the equations of motion for $SU(2)$ and $SU(3)$ center vortices in terms of a reduced set of profile functions, while in \S \ref{complex}
we discuss general properties of junctions and the associated non Abelian phases. In \S \ref{appa}, the natural choice 
of gauge group and quark sources in the YMH model is suggested. This is done in the framework of lattice inspired effective models,
which also deal with topological defects, but on the side of the pure YM theory we are intending to describe. Finally, in \S \ref{conc}
we present our conclusions.

\section{The Yang-Mills-Higgs model}
\label{YMHm}

As is well known, the $SU(2)$ YMH model with a Higgs field in the adjoint, in the phase where the symmetry is spontaneously broken down to $U(1)$, leads to the t' Hooft-Polyakov monopole. When a pair of adjoint Higgs fields is considered, the SSB phase only preserves $Z(2)$, and the model contains $Z(2)$ vortices  \cite{deVega}-\cite{HV}.  Motivated by the description of correlated singular $Z(2)$ vortices and monopoles in $SU(2)$ YM  in terms of a local color frame $n_1$, $n_2$, $n_3$ \cite{lucho}, it is expected that an  $SU(2)$ YMH model containing three adjoint Higgs fields will lead not only to smooth $Z(2)$ vortices but also to smooth junctions formed by a pair of vortices attached to a monopole-like configuration.

 In this chapter we shall start by defining this type of model, generalized to the case of a connected compact simple Lie algebra $\mathfrak{g}$.  The action is given by,
\begin{equation}
S= S_{\rm YM} + S_{\rm Higgs}\;,
\label{action}
\end{equation}
which contains the Yang-Mills term,
\begin{equation}
S_{\rm YM}= -\int d^4 x\; \frac{1}{4} \vec{F}_{\mu \nu}\cdot \vec{F}^{\mu \nu}\;,
\label{SYM}
\end{equation}
\begin{equation}
F_{\mu \nu} = \vec{F}_{\mu \nu} \cdot \vec{T}=(i/g)\left[D_\mu,D_\nu \right]
\makebox[.3in]{,}
D_\mu=\partial_\mu-ig A_\mu,
\end{equation}
where $A_\mu = \vec{A}_{\mu}\cdot \vec{T}$, and the $T_A$'s, $A=1, \dots, d$ are hermitian generators of 
$\mathfrak{g}$,
\begin{equation}
[T_A,T_B]=if_{ABC} T_{C}\;.
\label{Lie-alg}
\end{equation}
In the algebra, a positive definite metric exists,
\begin{equation}
\langle X ,Y\rangle =Tr \left(Ad(X) Ad(Y)\right)\footnote{If we were using a convention with antihermitian generators, in order to work with a positive definite metric, a minus sign should be included.}\;,
\end{equation}
where $Ad(X)$ is a linear map of $X\in \mathfrak{g}$ into the adjoint representation, generated by $d\times d$ hermitian matrices 
 $M_A$, with elements $M_A|_{BC}=-i f_{ABC}$, satisfying,
\begin{equation}
\left[M_{A},M_{B}\right]=if_{ABC} M_C\;,
\label{algeb}
\makebox[.5in]{,}
Tr(M_A M_B)=\delta_{AB}\;.
\label{alg-norm}
\end{equation}
With this normalization, we have, 
\begin{equation}
\langle T_A,T_B\rangle = \delta_{AB}
\makebox[.5in]{,}
f_{ABC}\, f_{DBC} =\delta_{AD} \;.
\label{norma}
\end{equation}

The Higgs sector is given by a set of fields $\psi_I = \vec{\psi}_I \cdot \vec{T}$, transforming in the adjoint,
\begin{equation}
\psi _I \rightarrow U \psi_I U^{-1}
\makebox[.5in]{,}
U \in G\;,
\label{Ht}
\end{equation}
and the Yang-Mills-Higgs action is gauge invariant under the {\it regular} transformation,
\begin{equation}
A_\mu \rightarrow A^U_\mu  = U A_\mu U^{-1} +\frac{i}{g} U \partial_\mu U^{-1}\;,
\label{At}
\end{equation}
when accompanied by local transformations (\ref{Ht}). That is,
\begin{equation}
S_{\rm Higgs}=\int d^4 x\, \left( \frac{1}{2} \langle D_\mu \psi_I , D^\mu \psi_I\rangle -
V_{\rm Higgs}(\psi_I)\right) \;,
\label{HiggsA}
\end{equation}
\begin{equation}
D_\mu \psi_I= \partial_\mu \psi_I -ig[A_\mu,\psi_I]\;.
\label{cov-abs}
\end{equation}

To construct a Higgs potential, defined on the fields $\psi_I=\vec{\psi}_I \cdot \vec{T}$, the natural invariant terms involve the metric applied to pairs of Lie algebra elements. Note also that as we are using hermitian generators, the closed
product in the algebra is in fact given by,
\begin{equation}
\psi_I \wedge \psi_J = -i [\psi_I,\psi_J]\;.
\end{equation}
Then, up to quartic order, the terms we shall consider are,
\begin{equation}
\langle \psi_I,\psi_J \rangle
\makebox[.3in]{,} \langle\psi_I,\psi_J \wedge \psi_K\rangle
\makebox[.3in]{,}  \langle \psi_I\wedge \psi_J,\psi_K \wedge \psi_L\rangle
\makebox[.3in]{,}
\langle \psi_I,\psi_J \rangle \langle \psi_K,\psi_L \rangle
\;.
\end{equation}
At this point, an additional symmetry will be introduced, acting on the flavor, with {\it global} group $Ad(G)$. For this aim, we note that the action of the gauge group in eq. (\ref{Ht}) can be represented using the components of $\psi_I$ in the Lie basis, 
\begin{equation}
\psi_I=\vec{\psi}_I|_A\, T_A \rightarrow U\psi_I U^{-1}=\vec{\psi}_I|_B\, U T_B U^{-1}\;.
\end{equation}
As the transformation of the $T_B$'s is a linear combination of generators, we can write,
\begin{equation}
U T_B U^{-1}= T_A\, R_{AB}\;,
\label{unr0}
\end{equation}
\begin{equation}
\vec{\psi}_I|_A \rightarrow R_{AB}\, \vec{\psi}_I|_B
\makebox[.5in]{,}
R = R (U)\in Ad(G)\;,
\label{unr}
\end{equation}
where $R$ is the $d\times d$ matrix that corresponds to the adjoint representation of the group.
Then, to implement the additional symmetry, the index $I$ will run from $1$ to $d$ and, renaming $I\to A$, we shall require the action to be invariant under global transformations acting on the flavor index,
\begin{equation}
\psi_A \to {\cal R}_{AB}\, \psi_B 
\makebox[.5in]{,}
{\cal R}\in Ad(G)\;.
\end{equation}
Up to quartic order, this restricts the possible terms in the potential to be \footnote{A $\langle \psi_A, \psi_B\rangle^2$ term could also be added, although the subsequent analysis would not be essentially modified.},
\begin{eqnarray}
\lefteqn{V_{\rm Higgs}= c+ \frac{\mu^2}{2}\, \langle \psi_A ,\psi_A \rangle+ } \nonumber \\
&&+\frac{\kappa}{3} \,f_{ABC} \langle \psi_A \wedge \psi_B,\psi_C \rangle +\frac{\lambda}{4}\, \langle \psi_A \wedge \psi_B,\psi_A\wedge \psi_B \rangle \;.
\label{Higgs-po}
\end{eqnarray}
Taking the first order variation of the potential with respect to $\psi_A \to \psi_A + \delta\psi_A$, the critical points are obtained from,
\begin{eqnarray}
 \mu^2\, \psi_A + \kappa \, f_{ABC}\, \psi_B \wedge \psi_C 
+\lambda\,  \psi_B \wedge (\psi_A \wedge \psi_B)  =0\;.
\label{critical}
\end{eqnarray} 
The possible solutions have the form,
\begin{equation}
\psi_A=v\, S T_A S^{-1}
\makebox[.5in]{,}
S\in G\;,
\label{form}
\end{equation}
with the condition,
\begin{equation}
\mu^2 v +\kappa v^2 + \lambda v^3 =0\;,
\end{equation}
that is, $v = 0$, or
\begin{equation}
v=v_c=-\frac{\kappa}{2\lambda}\pm \sqrt{\left(\frac{\kappa}{2\lambda}\right)^2-\frac{\mu^2}{\lambda }}\;.
\label{crit}
\end{equation}
Which configurations correspond to local or global minima depends on the choice of parameters. The associated Higgs potential is, 
\begin{equation}
V_{\rm Higgs}= c+ \frac{\lambda d}{4}\, v^2 \left[(v-v_0)^2+ b^ 2 \right] \;,
\end{equation}
\begin{equation}
v_0=-\frac{2\kappa}{3\lambda}
\makebox[.5in]{,}
b^2=\frac{2\mu^2}{\lambda }-\left(\frac{2\kappa}{3\lambda}\right)^2 \;,
\end{equation}
so in order to have an energy bounded from below, we must always have $\lambda > 0$. We also note that changing in the potential $\kappa\to -\kappa$ is equivalent to $v \to -v$,
so we can start analyzing the $\kappa < 0$ case, and then extend the result to positive values. 

For control parameters in the range $\mu^2>\frac{1}{4}\frac{\kappa^2}{\lambda} $, there is only one symmetric minimum at $\psi_A=0$. When 
$\frac{1}{4}\frac{\kappa^2}{\lambda}  > \mu^2 >\frac{2}{9}\frac{\kappa^2}{\lambda}$, the global minimum is still at $\psi_A=0$, and minima 
with higher energy appear at $\psi_A=v_c\, S T_A S^{-1}$, with $v_c >0$ computed using the positive root in eq. (\ref{crit}). At $\mu^2 =\frac{2}{9}\frac{\kappa^2}{\lambda}$, these minima also become ground 
states, degenerate with the symmetric one, and for $\mu^2 < \frac{2}{9}\frac{\kappa^2}{\lambda}$ their energy is lowered. In the SSB phase,  the space of global minima is labeled by the adjoint representation of the group,
\begin{equation}
{\cal M}=\{ \psi_A=v_c\, n_A\,,~~ n_A=S T_A S^{-1} ,~~ S\in G\}\;,
\label{vman-0}
\end{equation}
where the $n_A$'s form a local Lie basis,
\begin{equation}
[n_A,n_B]=if_{ABC}\, n_C
\makebox[.5in]{,}
\langle n_A , n_B \rangle =\delta_{AB}\;.
\label{localie}
\end{equation}
In components with respect to the basis $T_A$, we can also write,
\begin{equation}
{\cal M}=\{\vec{\psi}_A = v_c\,\hat{n}_A \,,~~ \hat{n}_A=R(S)\, \hat{e}_A\,,~~ R \in Ad(G)\}\;,
\label{vman}
\end{equation}
where $\hat{e}_A$ is a $d\times 1$ matrix, with its $A$-th element equal to $1$, and the rest equal to zero. 

When the system chooses a given vacuum, say $S=I$, the continuous $U$-symmetry of the potential is spontaneously broken to a discrete center symmetry $\psi_A \to U \psi_A U^{-1}$, $U \in Z(G)$, so all the components of the gauge 
field become massive.  In addition, the continuous global symmetry 
\begin{equation}
\psi_A \to  R_{AB}\,
 U \psi_B U^{-1}
\makebox[.5in]{,}
R=R(U)\;,
\end{equation}
that involves global gauge and flavor transformations, remains unbroken.

Note that unlike the regular gauge transformations $R(U)$, when dealing with topological objects, the extension of the $x$-dependent vacua to the whole space will be associated with a non Abelian phase $R(S)$ containing defects. These will correspond to the guiding centers of vortices and monopoles.

For $\kappa >0$, we have the same picture, with the minima at $v_c<0$ corresponding to the negative root in eq. (\ref{crit}), and at $\kappa=0$, $\mu^2<0$, there is a line where two types of degenerate ground states of the form $ \{ \psi^\pm_A = v_\pm\, ST_AS^{-1},~~  v_\pm = \pm \sqrt{-\mu^2/\lambda} \}$ coexist.

\subsection{Cartan decomposition of $\mathfrak{g}$}

In the next sections we shall discuss center vortices and junctions. For this aim, it will be useful considering the Cartan decomposition of the Lie algebra. 

A compact connected simple $\mathfrak{g}$ can be decomposed in terms of diagonal generators $T_q$, $q=1,\dots, r$,
which generate a Cartan subgroup $H$, and off-diagonal generators $E_{\alpha}$, or root vectors of the adjoint representation, labeled by the roots $\vec{\alpha}=(\alpha_1, \dots, \alpha_r)$,  satisfying,
\begin{equation}
[T_q,T_p]=0 \makebox[.3in]{,}
[T_q,E_\alpha]=\alpha_q\, E_\alpha
\makebox[.3in]{,}
[E_\alpha,E_{-\alpha}]=\alpha^q\, T_q\;.
\label{algebrag}
\end{equation}
\begin{equation}
 [E_\alpha, E_{\gamma}]=N_{\alpha \gamma}\, E_{\alpha+\gamma}
\makebox[.3in]{,}  \vec{\alpha}+\vec{\gamma} \neq 0 \;,
\end{equation}
where $N_{\alpha \gamma}=0$, if $\vec{\alpha}+\vec{\gamma}$ is not a root. The off-diagonal generators
can be grouped in pairs $E_{\alpha}, E_{-\alpha}$, where $\vec{\alpha}$ are the positive roots. The hermitian generators, can be identified with,
\begin{equation}
\{ T_q\} \makebox[.5in]{,}
\{ T_a \}=\left\{ T_{\alpha} ~,~
T_{\bar{\alpha}} \right\}\makebox[.5in]{,}\vec{\alpha}>0\;,
\end{equation}
\begin{equation}
T_{\alpha}=\frac{1}{\sqrt{2}}(E_\alpha + E_{-\alpha})
\makebox[.5in]{,}
T_{\bar{\alpha}}=\frac{1}{\sqrt{2}i}(E_\alpha - E_{-\alpha})  \;,
\label{Tes}
\end{equation}
where the following triplets generate $\mathfrak{so}(3)$ subalgebras,
\begin{equation}
\frac{1}{\alpha^2}\, \vec{\alpha}\cdot
\vec{T}\makebox[.5in]{,}
\frac{1}{\sqrt{\alpha^2}}\, T_{\alpha}
\makebox[.5in]{,}
\frac{1}{\sqrt{\alpha^2}}\, T_{\bar{\alpha}}\;.
\label{subalg}
\end{equation}

Note that for the set of all hermitian generators we are using upper-case letters, $T_A$, reserving
lower-case letters, in the beginning of the alphabet, for the off-diagonal sector, $T_a$.

The weights of a given matrix realization of the Lie algebra will also play an important role. A weight $\vec{w}$ is defined by the eigenvalues of diagonal generators corresponding to one common eigenvector.
In particular, the second equation in (\ref{algebrag}) says that the roots are the weights of the adjoint representation, which acts via commutators. We shall also refer to the eigenvectors as weight vectors.

\subsection{Local minima}

At the classical level, we shall be interested in looking for 
topologically stable (dual) magnetic static solutions to the equations of motion, whose total energy is given by, 
\begin{equation}
E= \int d^{3} x\, \left(\frac{1}{4} \langle F_{i j}, F_{i j}\rangle + \frac{1}{2} \langle D_i \psi_A , D_i \psi_A \rangle +
V_{\rm Higgs}(\psi_A)\right)\;.
\label{energy}
\end{equation}
We are particularly interested in the SSB phase, where the global minima are given by $Ad(G)$. In this phase, let us intially analyze the local ground states, that is, the general conditions necessary to have, in a given asymptotic region, energy densities tending to zero. 
When discussing static configurations, we shall partially fix the gauge by the condition $A_0=0$. More generally, we could also be interested in analyzing the path integral quantization of models containing a Yang-Mills term in $4D$ Euclidean spacetime, where relevant configurations must have finite action. In that case, the asymptotic conditions on the gauge field $A_\mu$ are simply obtained by replacing 
$i \to \mu$ in the equations below.

In the asymptotic region, not only the condition (\ref{vman-0}) but also $D_i \psi_A = 0$ must be imposed,
\begin{equation}
D_i n_A = 0
\makebox[.3in]{,}
n_A=ST_A S^{-1}
\;.
\label{stronger}
\end{equation}
The most general $A_i $ can be {\it locally} written as $A_i\,|_{\rm loc}={\cal A}^S_i$, a gauge transformation satisfying
$D_i({\cal A})\, T_A = 0$, that is,
\begin{equation}
[{\cal A}_i, T_A]=0\;.
\label{zerocov}
\end{equation}
As this is valid for every $A=1,\dots, d$, and the algebra is simple, this implies, 
\begin{equation}
{\cal A}_i=0
\makebox[.5in]{,}
A_i\, |_{\rm loc} = \frac{i}{g} S \partial_i S^{-1}\;,
\label{suncoc}
\end{equation}
leading to $F_{i j}=0$ on that region.

We would like to underline that in general, even in the asymptotic region, eq. (\ref{suncoc}) is only valid locally, as $S$ could change by a center element when we go around a loop. How to extend this expression to the whole asymptotic region depends on the global structure of $G$ used to write the model. However, as all the fields are in the adjoint, any structure represents the same model.
One possibility, when using a general representation $S$, is to write the asymptotic configuration in the global form,
\begin{equation}
A_i = \frac{i}{g} S \partial_i S^{-1} -I_i\;,
\end{equation}
where $I_i$ is concentrated on the surfaces where $S$ changes by a center element, and is designed to compensate
the derivative of the discontinuities in the first term. This type of approach has been introduced in \cite{engelhardt1,reinhardt} in the context of pure YM theories. This is simplified when using the adjoint representation of the group, 
which does not ``see'' the center of the group,
\begin{equation}
A^A_i\, M_A = \frac{i}{g} R \partial_i R^{-1}
\makebox[.5in]{,}
R=R(S)\;,
\label{rcoc}
\end{equation}
as $R$ will always be single-valued when we go around a loop.  Another convenient manner to deal with this problem, in any representation, is by writting the asymptotic gauge field in terms
of the single-valued quantities $n_A=ST_AS^{-1}$ \cite{lucho}, \cite{Lucho1}-\cite{Lucho3}.
In this respect, using that in the asymptotic region $A_i$ is defined by eq. (\ref{stronger}), with $A=1,\dots,d$, we have,
\begin{equation}
[n_A, D_i n_A]=0\;,
\end{equation}
and using the property,
\begin{equation}
n_A \wedge (X \wedge n_A) 
= \langle X , n_B\rangle\, n_A \wedge (n_B \wedge n_A) = X\;,
\end{equation}
we get,
\begin{equation}
A_i = -(1/g)\, n_A \wedge \partial_i n_A \;.
\label{rep1}
\end{equation}
Alternatively, expanding $\partial_i n_A= \langle n_B,\partial_i n_A \rangle \, n_B$, we can also write,
\begin{eqnarray}
A_i = -C^A_i\, n_A 
\makebox[.5in]{,}
C^A_i = -(1/g)\,f_{ABC}\, \langle n_B,\partial_i n_C \rangle\;
\label{rep2}
\end{eqnarray}
(see also ref. \cite{Lucho2}).

This can be extended to the whole space, so that the gauge field $A_i$, whose general form is locally given by, 
\begin{equation} 
A_i \,|_{\rm loc} =S\, {\cal A}_i S^{-1} +\frac{i}{g} S\partial_i S^{-1}\;,
\label{Agen}
\end{equation}
can be globally written in the forms, 
\begin{equation} 
A^A_i M_A = R\, ({\cal A}^A_i M_A) R^{-1} +\frac{i}{g} R\partial_i R^{-1}
\makebox[.5in]{,}
R=R(S)\;,
\label{AR}
\end{equation}
\begin{equation}
A_i = ({\cal A}^A_i-C^A_i)\, n_A  \makebox[.5in]{,}
{\cal A}_i= {\cal A}^A_i\, T_A\;,
\label{ansatz}
\end{equation}
with the condition,
\begin{equation}
{\cal A}^A_i \to 0\;,
\label{localm}
\end{equation}
when approaching the asymptotic region. At the guiding centers of the topological objects, some frame components $n_A$ will contain defects and, as a consequence, some of the frame dependent fields $C^A_i$ will be singular. For this reason, we shall need to impose additional boundary conditions on the gauge and Higgs fields. In fact, we will see in \S \ref{wc}, \S \ref{cva} and \S \ref{complex} that among the quantities $C^A_i$, some diagonal components play a similar role to $\partial_i \varphi$,  when analyzing a Nielsen-Olesen vortex, while some combinations of the off-diagonal components, a similar role to 
$\hat{r}\times \partial_i \hat{r}$, when analyzing a 't Hooft-Polyakov monopole \cite{Manton}.

To compute the field strength, we can use eq. (\ref{AR}) and recall that the commutator $[\partial_i, \partial_j]$ is nonzero when applied on singular quantities, to obtain,
\begin{eqnarray}
F^A_{i j} M_A
&=& R (F^A_{i j}({\cal A})M_A) R^{-1}
+(i/g) R[\partial_i,\partial_j]R^{-1}\;,
\label{Fcalas} 
\end{eqnarray} 
where the second term can also be rewritten as $-(i/g)R(R^{-1}[\partial_i,\partial_j]R) R^{-1}$.
In addition, from eq. (\ref{rep2}), we have,
\begin{equation}
C^A_i\, M_A =\frac{i}{g} R^{-1} \partial_i  R \;,
\makebox[.5in]{,}
 (i/g) R^{-1} [\partial_i,\partial_j]  R=  F^A_{i j}(C)\, M_A \;.
\label{CAm}
\end{equation}
Then, putting these pieces together, we get,
\begin{equation}
F_{i j}=F^A_{ij}\, T_A = G^A_{i j}\, n_A
\makebox[.5in]{,}
G^A_{i j}= F^A_{i j}({\cal A}) - F^A_{i j}(C)\;,
\label{field-st} 
\end{equation}
where $F^A_{i j}(C)$ is concentrated at the frame defects \cite{Lucho2},
\begin{eqnarray}
F^A_{i j}(C) &=& (i/g)\, {\rm tr}\, (M^A
R^{-1}[\partial_i,\partial_j]R) \nonumber \\
&=&\partial_i C^A_j-\partial_j C^A_i + g f^{ABC} C^B_i C^C_j\;.
\label{FCe}
\end{eqnarray}

\subsection{The Yang-Mills-Higgs equations}

The minimization of the energy (\ref{energy}) gives,
\begin{equation}
 D_j F_{ij} = ig\, [\psi_A,D_i\psi_A]\;,
\label{YMF}
\end{equation}
\begin{equation}
 D_i D_i \psi_A = \mu^2 \psi_A + \kappa\, f_{ABC}\, \psi_B \wedge \psi_C +\lambda\, \psi_B \wedge (\psi_A \wedge \psi_B)\;.
\label{YMP}
\end{equation}
For a given map $S$, defining a local frame $n_A$ containing defects, we can look for solutions with the gauge fields 
given by eq. (\ref{ansatz}) and the Higgs fields by,
\begin{equation}
\psi_A = S (h_{AB}\, T_B) S^{-1}\;,
\label{hyS}
\end{equation}
where $h_{AB}$ are real symmetric smooth profile functions.

Using eq. (\ref{Agen}), or its global form (\ref{AR}), at every point but on the zero measure region of frame defects, 
we get,
\begin{equation}
 {\cal D}^{AB}_j F^B_{ij}({\cal A}) = g\, f_{ABC}\, h_{DB} \,D^{CE}_ih_{DE}
\makebox[.5in]{,}
{\cal D}^{AB}_i=\delta_{AB}\, \partial_i-g\, f_{ABC} \, {\cal A}^C_i
\;,
\label{YM1}
\end{equation}
\begin{eqnarray}
{\cal D}^{BC}_i {\cal D}^{CD}_i h_{AD} = \mu^2 \, h_{AB}+ \kappa\, f_{BCD} f_{AEF}\, h_{EC} h_{FD}&&\nonumber \\ 
+ \lambda\,  f_{BCD} f_{DEF}\, h_{GC} h_{AE} h_{GF}\;.&&
\label{YM2}
\end{eqnarray}
In the asymptotic region, besides ${\cal A}_i \to 0$, we have to impose,
\begin{equation}
h_{AB} \to v_c\, \delta_{AB}\;,
\end{equation}
and, at the guiding centers, $h_{AB}$ and  ${\cal A}_i$ must satisfy boundary conditions such that $A_i$ and $\psi_A$, in eqs. (\ref{ansatz}) and (\ref{hyS}), be smooth. These requirements will be further modified by the addition of external quarks sources. 

In \S \ref{cva}, we shall give a center vortex ansatz such that the equations of motion can be written in terms of a single profile function for the gauge field components, and a reduced set of Higgs profiles. Before doing this, general model independent properties about the topology, weights and charges of vortices and junctions are developed in the following sections. 

\section{The topology of center vortices}
\label{topo-cent}

In the absence of external quark sources, the relevant global structure of the model is that of the adjoint. It can be given by $Ad(G)=G/Z(G)$, where $Z(G)$, the center of $G$, is the kernel of the homomorphism $G\to Ad(G)$. This is a discrete subgroup formed by those elements $S\in G$ verifying $ST_AS^{-1}=T_A$, for $A=1,\dots , d$. 

In particular, to compute the fundamental homotopy group of the adjoint, it is useful considering the universal covering group $\tilde{G}$, uniquely defined as the simply connected group whose Lie algebra is $\mathfrak{g}$, and apply the exact homotopy sequence,
\begin{equation}
0=\Pi_1 ( \tilde{G})\rightarrow \Pi_1 (\tilde{G}/Z(\tilde{G}))\rightarrow \Pi_0(Z(\tilde{G})) \rightarrow \Pi_0(\tilde{G})=0\;,
\label{sequence}
\end{equation}
to conclude that the fundamental homotopy group of the space of vacua ${\cal M}$ in eqs. (\ref{vman-0}), (\ref{vman}) is,
\begin{equation}
\Pi_1({\cal M}) = \Pi_1(Ad(G))=Z(\tilde{G})\;.
\label{fhom}
\end{equation}

For example, when $G=SU(N)$, $Z(\tilde{G})=Z(N)$.  When we go along an open path $S(\tau)\in SU(N)$, $\tau \in [0,1]$, that changes by a center element $e^{i2\pi/N}$, a nontrivial closed path in the adjoint representation of SU(N) is obtained. If this path is composed $N$ times we get a trivial map in the adjoint, as it is associated with a closed path in the fundamental representation of SU(N), whose first homotopy group is trivial. This 
is the meaning of $\Pi_1(Ad(SU(N)))=Z(N)$ in eq. (\ref{fhom}). 

For a straight center vortex, the asymptotic region is given by $R^3-{\rm T}^2$, the complement of a cylinder ${\rm T}^2$ containing the vortex.  The defining property of a center vortex is that for any closed path contained in the asymptotic region, linking the vortex core $T^2$, the Wilson loop,
\begin{equation}
W[A]= (1/d)\,tr\, P \exp \left(ig\oint dx_\mu\, A_\mu\right)\;,
\label{wilson}
\end{equation} 
computed using the group elements in $\tilde{G}$, gives a center element in $Z(\tilde{G})$ ($P$ stands for path ordering). Note that this is a gauge invariant definition, as for regular gauge transformations $W[A^U]=W[A]$. We would also like to underline that it is only in the context of pure Yang-Mills theories, where the gauge field represents the gluons and the path a quark/antiquark pair, that the Wilson loop is a direct measure of the interquark potential. In the dual YMH model, in order to represent the quarks, appropriate external pointlike monopole sources will have to be introduced, see the discussion in \S \ref{appa}.

Now, as in the asymptotic region the Wilson loop assumes a discrete value (the group is simple) it cannot be changed by continuous deformations of the smooth localized ansatz, as long as the loop be always chosen contained in the asymptotic region of local minima. Then, when $\Pi_1(Ad(G))$ is nontrivial, there will be static center vortex configurations with finite energy per unit length of the vortex. 

\subsection{Center vortex weights}
\label{wc}

Let us consider a vortex characterized by a center element $\mathfrak{z}\in Z(\tilde{G})$. This means that the asymptotic gauge 
field must be such that,
\begin{equation}
W[A]=\mathfrak{z} \;,
\label{Wzeta}
\end{equation}
where ${\cal C}$ is a loop completely contained in the asymptotic region, linking the vortex core. This is the case when
using in eqs. (\ref{suncoc}), (\ref{rcoc}),
\begin{equation}
S= e^{i\varphi\, \vec{\beta} \cdot \vec{T}}
\makebox[.5in]{,}
R= e^{i\varphi\, \vec{\beta}\cdot \vec{M}}
\;,
\label{RSvort}
\end{equation}
with the possible $\vec{\beta}$'s satisfying,
\begin{equation}
e^{i2\pi\, \vec{\beta} \cdot \vec{T}} = \mathfrak{z}\, I \in Z(\tilde{G})\;,
\label{poss-b}
\end{equation}
thus leading to a single-valued $R$ transformation in $Ad(G)$,
\begin{equation}
R(2\pi)= e^ {i2\pi\, \vec{\beta} \cdot \vec{M}}=I_{d\times d}
\;.
\label{pweights}
\end{equation}
Here, $\varphi$ is the polar angle with respect to the center vortex axis, and 
we adhered to the convention that in dot products involving a weight, $\vec{\beta}$ is considered as
$(\beta_1,\dots, \beta_r,0,\dots,0)$.

The possible center vortex weights have been discussed in refs. \cite{Fidel2,Konishi-Spanu}, following the ideas introduced in \cite{GNO} to characterize non Abelian monopoles.
Using eq. (\ref{pweights}) on the weight vectors of the adjoint representation $Ad(G)$, we get,
\begin{equation}
e^ {i2\pi\, \vec{\beta} \cdot \vec{\alpha}}=1,
\label{sing-v}
\end{equation}
where $\vec{\alpha}$ is a weight of the adjoint, that is, a root of $\mathfrak{g}$. Then, for every $\vec{\alpha}$, the possible weights must satisfy the condition,
\begin{equation}
\vec{\beta} \cdot \vec{\alpha}\in Z\;.
\label{qc}
\end{equation}
whose solution is given by  \cite{GNO}
\begin{equation}
\frac{1}{2\bar{N}}\, \vec{\beta} \in \Lambda(\tilde{G^{\sf v}})\;.
\label{solution}
\end{equation}
That is, when the Higgs fields transform in $Ad(G)=\tilde{G}/Z(\tilde{G})$, the possible $\vec{\beta}$'s are associated with the weights of the universal covering of the dual group $G^{\sf v}$, whose Lie algebra is 
defined by the dual root system,
\begin{equation}
\vec{\alpha}^{\,\sf v} = \frac{1}{\bar{N}}\frac{\vec{\alpha}}{\alpha^2}\;.
\end{equation}
For simple Lie algebras, $\bar{N}$ is a number, or rescaling parameter, that only depends on the general structure of $\mathfrak{g}$.  
In the lattice $\Lambda(\tilde{G^{\sf v}})$, there are many types of center vortices labeled by $\mathfrak{z}$, although this value does not fix uniquely the asymptotic behavior. A pair of asymptotic non Abelian phases defined by $\vec{\beta}$ and $\vec{\beta}'$
are topologically equivalent if the map
\begin{equation}
R R'^{-1} = e^{i\varphi\, (\vec{\beta}-\vec{\beta}')\cdot \vec{M}},
\label{equiv-cond}
\end{equation}
can be continuously deformed to the trivial one. This is the situation when the associated path in the simply connected $\tilde{G}$
is closed. Then, substituting $\vec{M}$ in eq. (\ref{equiv-cond}) by generators of the universal covering $\tilde{G}$, and acting on the possible weight vectors, two vortices are topologically equivalent if,
\begin{equation}
e^ {i2\pi\, (\vec{\beta}-\vec{\beta}') \cdot \vec{w}}=1\;,
\label{qe}
\end{equation}
where $\vec{w}$ is any weight of the universal covering $\tilde{G}$. This means that the closed path in $\Pi_1(Ad(G))$ will also be closed in $\tilde{G}$ that, being simply connected, will permit the continuous deformation of the composition (\ref{equiv-cond})  to a point. The solution to eq. (\ref{qe}), i.e., $(\vec{\beta}-\vec{\beta}') \cdot \vec{w} \in Z$ is given by,
\begin{equation}
\frac{1}{2\bar{N}}\, (\vec{\beta}-\vec{\beta}') \in \Lambda(Ad(G^{\sf v}))\;,
\label{equiv-vc}
\end{equation}
the lattice of the dual root system \cite{GNO}. 

In other words, a center vortex defined by $\mathfrak{z}$ can be associated with equivalence classes of asymptotic behaviors, 
\begin{equation}
Z(\tilde{G})=\Lambda(\tilde{G^{\sf v}})/\Lambda(Ad(G^{\sf v}))\;.
\end{equation}
The weights of a given irreducible representation only belong to one equivalence class \cite{Brian}. 

An important point we would like to emphazise is that, for a given $\mathfrak{z}$, we are interested in determining the smooth configuration that minimizes the energy. If we start with an asymptotic behavior labeled by $\vec{\beta}'$, the search for this configuration should not only contemplate the direct extension of the phase inside the core, keeping it along a Cartan direction, but also other possibilities. The behavior $S= e^{i\varphi\, \vec{\beta}' \cdot \vec{T}}$ on a circle contained in the asymptotic region could also be continuously changed to $S= e^{i\varphi\, \vec{\beta}\cdot \vec{T}}$, with $\vec{\beta}\neq \vec{\beta}'$ satisfying eq. (\ref{equiv-vc}), when approaching the vortex axis. In that case, the extended phase $S(x)$ would use the whole non Abelian group. 

For example, an asymptotic behavior labeled by a nonzero weight in $ \Lambda(Ad(G^{\sf v}))$ can be continuously converted into the behavior $S\sim I$ at the $z$-axis. 
Then, in this case, there are in fact no defects, and the minimization process will simply return a trivial result, corresponding to a pure gauge transformation of the configuration $A_\mu \equiv 0$, $\psi_A \equiv v_c\, T_A $. 
On the other hand, when the asymptotic weight is not in  $ \Lambda(Ad(G^{\sf v}))$, any phase extension inside the vortex will necessarily lead to a local frame $n_A=ST_A S^{-1}$ containing defects.  For those $\vec{\beta}$ values characterizing the behavior  of the stable solutions at the origin, it will be possible to perform a gauge transformation such that the phase is everywhere along the Cartan sector. In that case, it will be useful to compute the Abelian charge of the center vortex. For the minimum charge center vortices in $SU(N)$, these values are expected to be the weights of the fundamental representation \cite{Konishi-Spanu}. 

\subsection{Charges of center vortices}
\label{charge-v}

When the smooth configuration minimizing the energy contains the Abelian phase factor (\ref{RSvort})
at every point $x\in R^3$, the local frame $n_A$, the frame dependent field $C^A_i$ in eq. (\ref{rep2}), and  $F^A_{i j}(C)$ can be easily computed.  For example, using eqs.  (\ref{CAm}), (\ref{FCe}),  the nonzero components are the diagonal ones, $A=q$,
\begin{equation}
C^q_i = -\frac{1}{g} \, \partial_i \varphi\, \vec{\beta}|^q
\makebox[.5in]{,}
F^q_{i j}(C) = -\frac{1}{g} \, [\partial_i,\partial_j] \varphi\, \vec{\beta}|^q
\label{F-cv}
\;.
\end{equation}
In that case, in order for the gauge field and field strength be well defined smooth functions, the condition,
\begin{equation}
{\cal A}_i^q \to C^q_i\;,
\end{equation}
must be satisfied at the center vortex axis. In addition, the off-diagonal components that rotate when we go around the
$z$-axis,
\begin{eqnarray}
n_\alpha =S T_\alpha S^{-1}
\makebox[.3in]{,}
n_{\bar{\alpha}}=S T_{\bar{\alpha}} S^{-1}
\makebox[.5in]{,}
\vec{\beta}\cdot \vec{\alpha}  \neq 0
\;,
\end{eqnarray}
contain a defect at the origin,
so those profile functions multiplying them should tend to zero there.
In Fig. 1, we show the local frame for a unit charge center vortex, in the case of $SU(2)$.

\begin{figure}
\centering
\includegraphics[scale=.4, bb = 0 0 350 350]{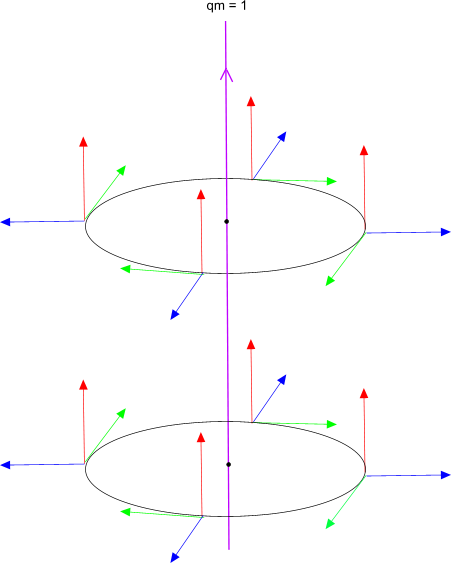}
\caption{A $Z(2)$ vortex. The red arrows represent the diagonal direction, while the green and blue arrows, the off-diagonal local ones.}
\end{figure}

The center vortex charges have been obtained in \cite{Konishi-Spanu}. Considering
the dual tensor components along the local direction $n_A$,
\begin{equation}
{\cal G}^A_i = \frac{1}{2}\, \epsilon_{ijk} G^A_{jk}\;,
\label{Bfield}
\end{equation}
and organizing the fluxes $Q_v^q = \int dS_i\, {\cal G}^q_i$ as
an $r$-tuple, $\vec{Q}_v=(Q^1_v,\dots, Q^r_v)$, 
we get,
\begin{equation}
\vec{Q}_v = \frac{2\pi}{g}\vec{\beta} \;.
\label{cv-ch}
\end{equation}
These charges are invariant under regular gauge transformations (see the discussion in \S \ref{cj}).

\section{The topology of junctions}
\label{chains-comp}

As the second homotopy group of a compact group is trivial, we have $$\Pi_2({\cal M})=\Pi_2(Ad(G))=0\;,$$ that is, there are no isolated monopole configurations.
However, we will see that center vortices can concatenate some monopole-like objects to form junctions. These will be nontrivial topological configurations along a local Cartan sector of the Lie algebra, necessary to interpolate the center vortices originated by SSB. In this sense, it is more appropriate referring to them as monopole-like junctions, as their existence depends on the attached vortices.

While a single center vortex is associated with smooth extended local Cartan directions $\hat{n}_q(x)=R(x)\, \hat{e}_q$, this will not be the situation in a monopole-like junction.
In order to understand them, let us first discuss the $\mathfrak{su}(2)$ case, and then generalize to $\mathfrak{g}$. 
In the framework of pure YM theories, we have seen in refs. \cite{lucho} that when using field decompositions involving a local color frame $\hat{n}_A$, $A=1,2,3$, chains formed by correlated monopoles and thin center vortices can be represented by thin junctions formed by correlated  defects. In $\mathfrak{su}(2)$ there is one Cartan generator $T_1$, and given a matrix $R\in Ad(SU(2))=SU(2)/Z(2)=SO(3)$, an associated local direction $\hat{n}_1=R\, \hat{e}_1$. On the other hand, 
given a unit vector $\hat{n}_1$, $R$ is defined up to right multiplication by a matrix leaving $\hat{e}_1$ invariant. 

Now, let us suppose we give 
a well defined configuration $\hat{n}_1(x)$, with $x$ on a surface $S^2$ around a given point. 
There are two possibilities, the first one is when this configuration can be generated by means of a well defined $R(x)$, $\hat{n}_1(x)= R(x)\, \hat{e}_1$, $x\in S^2$.
In that case, as $\Pi_2(SO(3))=0$, $R(x)$ can be continuously deformed to the trivial map $R\equiv I$, and in this process the configuration $\hat{n}_1(x)$ will be continuously deformed to the trivial one $\hat{n}_1(x)\equiv \hat{e}_1$. However, there is the possibility that a well defined configuration $\hat{n}_1(x)$ on $S^2$ be generated by an ill-defined $R(x)$. In such a case, we could not use the argument that $\Pi_2(SO(3))$ is trivial, to imply a trivial $\hat{n}_1(x)$, as $R(x)$ would not be defined on $S^2$.
Take for instance the configuration shown in Fig. 2, where, we have a hedgehog form for $\hat{n}_1(x)$ (red arrows), well defined on $S^2$. This is possible as there are two points on the sphere where the off-diagonal directions (green and blue arrows) are ill-defined. When we go around a small loop around the north (south) pole, $\hat{n}_a=R\, \hat{e}_a$, $a=2,3$ rotate once, so they are ill-defined at the poles. These rotations have the same orientation with respect to the outward normal, but different orientations with respect to $\hat{z}$. 

The presence of singularities can be understood in terms of the hairy ball theorem. When $\hat{n}_1$ is along the radial direction, $\hat{n}_2$ and $\hat{n}_3$ are tangent to $S^2$, so these fields must be ill-defined at a pair of points, namely, the guiding centers of the vortices. The non Abelian phase we have described then corresponds to a pair of center vortices, attached to a nontrivial configuration in the coset $SU(2)/U(1)=S^2=\{\hat{n}_1 \}$. Of course, to extend this type of local frame to a well defined field configuration on the three dimensional space, we still have to impose appropriate boundary conditions. On the $z$-axis, the profile functions multiplying $\hat{n}_2$ and $\hat{n}_3$ should tend to zero. Then,  when looking at the regular fields $\vec{\psi}_A$ on the sphere $S^2$, they will only form a basis at points other than the guiding centers of the vortices, this is the reason why these 
configurations are not governed by $\Pi_2(SO(3))=0$. In addition, the profile functions accompanying  $\hat{n}_1$ should tend to zero at the origin, where the hedgehog $\hat{n}_1$ is ill-defined. 
\begin{figure}
\centering
\includegraphics[scale=.4, bb = 0 0 350 350]{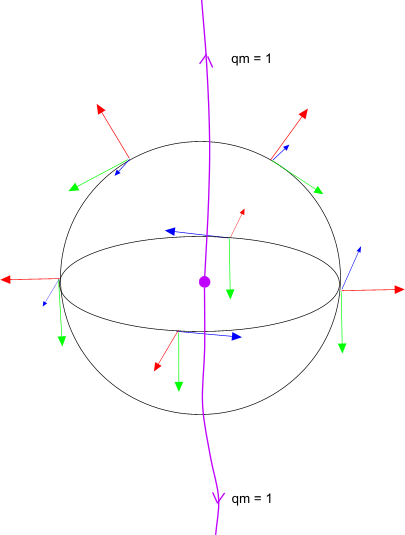}
\caption{A point-like monopole at the border of a pair of open $Z(2)$ vortices.}
\end{figure}

For general $G$, to construct the junction, we can start by giving a well defined  configuration for the local Cartan directions $n_q$, on a two-dimensional surface $S^ 2$ around a point. Again, when representing this configuration by $n_q=S T_q S^ {-1}$, or in components with respect to $T_A$ (see eq. (\ref{vman})),
\begin{equation}
\hat{n}_q(x)= R(x)\, \hat{e}_q
\makebox[.5in]{,}
x\in S^2 
\makebox[.5in]{,}
R \in Ad(G)\;,
\label{fR}
\end{equation}
$R(x)$ could be a well defined map on $S^2$, and in that case $\hat{n}_q(x)$ would be trivial, or  it could be ill-defined. That is, on some disks $D_{\rm c}\in S^2$ of small radius $\epsilon$, centered at points ${\rm c}={\rm c}_1, {\rm c}_2, \dots$,  $S(x)$ and $R(x)$ could behave as,
\begin{equation}
S_{{\rm c}}(x) \sim  {\cal S}_{{\rm c}} \,  e^{i\chi_{\rm c}\,\vec{\beta}_{\rm c}\cdot \vec{T}}
\makebox[.5in]{,}
R_{{\rm c}}(x)  \sim {\cal R}_{{\rm c}}\,   e^{i\chi_{\rm c}\,\vec{\beta}_{\rm c}\cdot \vec{M}}
\label{R-disk}
\end{equation}
(no summation over repeated c-type indices). Here, the first factors are smooth, while the angles $\chi_{\rm c}$ parametrize the borders $\partial D_{\rm c}$, oriented in the positive sense with respect to the outward normal to $S^2$. The $\vec{\beta}_{\rm c}$'s are in $\Lambda(\tilde{G^{\sf v}})$, the set of  possible weights leading to a single-valued $R_{\rm c}(x) $ when  we go around these borders (see \S \ref{wc}). With the behavior (\ref{R-disk}), the smoothness of  $\hat{n}_q(x)$ is guaranteed as the singular phases have no effect on $n_q$ at ${\rm c}$,
\begin{equation}
\lim_{x\to {\rm c}} n_q = {\cal S}_{\rm c} T_q {{\cal S}^{-1}_{\rm c}} 
\makebox[.5in]{,}
{\rm c}={\rm c}_1,{\rm c}_2,\dots\;.
\end{equation}
On the other hand, 
some off-diagonal components $\hat{n}_a(x)$ will be ill-defined at the points ${\rm c}$, which correspond to the guiding centers of the vortices attached to the monopole. 

An important global 
constraint on the possible field configurations is obtained by noting that, once the singular points are determined, the composition $$\Gamma=\gamma_1 \circ  \partial D_{{\rm c}_1}\circ \gamma_1^{-1} \circ \gamma_2 \circ  \partial D_{{\rm c}_2}\circ \gamma_2^{-1}\circ \dots \;,$$ is a closed path that can be deformed to a point in
$
S^2-(D_{{\rm c}_1}\cup D_{{\rm c}_2} \cup \dots)
$, a region free of singularities. Here, $\gamma_1, \gamma_2, \dots ,$ are open paths on $S^2$ joining a common point $x_0$ with points 
at $\partial D_{{\rm c}_1},
\partial D_{{\rm c}_2},\dots $, respectively. 
Let us suppose the above mentioned region is contained in the asymptotic region, that is, the field $A_\mu$ is pure gauge there.
Then, the Wilson loop computed in $\tilde{G}$, for any closed path contained in $S^2-(D_{{\rm c}_1}\cup D_{{\rm c}_2} \cup \dots)$, must give a center element in $Z(\tilde{G})$. Now, the center of a simple group is discrete, so that the value on $\Gamma$, 
\begin{equation}
W_{\Gamma}[A]= (1/d)\, tr\, [e^ {i2\pi\, \vec{\beta}_{{\rm c}_1} \cdot \vec{T}}\, e^ {i2\pi\, \vec{\beta}_{{\rm c}_2} \cdot \vec{T}} \dots] \;,
\end{equation}
cannot change as $\Gamma$ is deformed  to a point, using paths contained on a region free of singularities. That is, we obtain the constraint,
\begin{equation}
e^ {i2\pi\, \vec{\beta}_{{\rm c}_1} \cdot \vec{T}}\, e^ {i2\pi\, \vec{\beta}_{{\rm c}_2} \cdot \vec{T}} \dots = I
\makebox[.8in]{\rm or,}
\label{closedu}
\end{equation}
\begin{equation}
\frac{1}{2\bar{N}}\, (\vec{\beta}_{{\rm c}_1}+\vec{\beta}_{{\rm c}_2}+\dots) \in \Lambda(Ad(G^{\sf v}))\;.
\label{possible}
\end{equation}
This is a matching condition between the center vortices and the attached monopole.
In other words, the composition of the loops $e^{i\chi_{\rm c} \, \vec{\beta}_{\rm c}\cdot \vec{M}}$, ${\rm c}={\rm c}_1,{\rm c}_2,\dots,$ is an element in $Ad(H)$ that must be closed in $\tilde{G}$ and, therefore, topologically trivial in $Ad(G)$. 

The use of exact homotopy sequences is the appropriate framework to settle the discussion above, and obtain the 
conditions to have a topologically nontrivial vortex-monopole junction. These sequences have also been used to analyze non Abelian monopoles and vortices in systems with hierarchical gauge symmetry breaking \cite{notes}. The essential ingredient in our construction is that at any given point, the $n_q$'s define $R\in Ad(G)$ up to right multiplication by an element in $Ad(H)$. Then, the set of $n_q$'s can be identified with the quotient space $Ad(G))/Ad(H)$. 

As is well known, for relative homotopy classes associated with a pair of manifolds $X$, $Y$ 
there is an exact sequence:
\begin{equation}
\Pi_2(Y)\rightarrow \Pi_2(Y,X)\rightarrow \Pi_1(X) \rightarrow \Pi_1(Y)\;,
\end{equation}
such that the image of a given map in the sequence is  the kernel of the next map.
This theorem can be applied to $Y=Ad\, (G)$ and $X=Ad\, (H)$, to obtain,
\begin{equation}
0=\Pi_2(Ad(G))\rightarrow \Pi_2 \left(\frac{Ad\, (G)}{Ad\, (H)}\right)\rightarrow \Pi_1(Ad\, (H)) \rightarrow \Pi_1(Ad\, (G))=Z(\tilde{G})\;.
\label{g-sequence}
\end{equation}
The image of the first set in the sequence can only be a trivial map in the quotient, which coincides with the kernel of the second map.  In other words, different equivalence classes in $\Pi_2(Y/X)$ must correspond to different classes in $\Pi_1(X)$.  If the last 
set in the sequence were trivial, then the kernel of the third map, and therefore the image of the second, would be the whole $\Pi_1(X)$. In that case $\Pi_2(Y/X)$ would be homomorphic to $\Pi_1(X)$. However this is not the case, there are classes in $\Pi_1(X)$ that cannot be attained by the second map. The only classes attained are those mapped into the trivial class in $\Pi_1(Y)=Z(\tilde{G})$. 

The $S^1$ manifold in $\Pi_1(Y)$ precisely corresponds to $\partial D_{{\rm c}_1}\circ \partial D_{{\rm c}_2} \circ \dots$,
the closed path obtained by composing the closed paths around the center vortex defects. 
In this regard, note that in the simply connected case, the construction of the sequence starts by showing that any map $G(s,t)$, from a square into $G$, is equivalent to a map where three of the sides are mapped to the identity $I$, while the fourth is mapped into $H$. Identifying the sides adjacent to the fourth, we get a map from $D-D_1$, a disk with a hole at $D_1 \subset D$, with the external border $\partial D$ mapped into $I$, and the internal border $\partial D_1$ into $H$. As the points in $\partial D_1$ and $\partial D$ can be identified in $G/H$, an element in $\Pi_2(G/H)$ becomes correlated with an element in $\Pi_1(H)$. Note that in such configuration, the loop in $\Pi_1(H)$ must be trivial. When we continuously move from $\partial D_1$ to $\partial D$, it must be continuously deformed to the identity map. As $G$ is simply connected, this condition is automatically met.

However, when $Y$ is not simply connected, for a loop contained in $H$ at $\partial D_1$, that is nontrivial in $G$, this construction is not possible. For $Y=Ad(G)$, the best we can do is considering a disk $D$ with a number of holes at $D_1, D_2, \dots$,
where the external border $\partial D$ is mapped into $I$, and on the borders $\partial D_1, \partial D_2,\dots$ we have maps into $Ad(H)$,
with the condition that their composition be trivial in $Ad(G)$. This is the previously discussed junction composed by center vortices, with trivial total $Z(\tilde{G})$ charge, correlated with a monopole-like configuration in the coset 
$Ad(G)/Ad(H)=\{n_q\}$. 

When restricting to those configurations satisfying the matching condition, we have 
\begin{equation}
\Pi_2(Ad(G)/Ad(H)) = \left. \Pi_1(Ad(H))\right|_{\rm rest}  \;.
\end{equation}
Therefore, inequivalent elements in $\Pi_1(Ad(H))$ correspond to inequivalent elements in $\Pi_2(Ad(G)/Ad(H))$. Then, among the possible
charges in eq. (\ref{possible}), the condition for having a nontrivial monopole-like junction is 
\begin{equation}
\vec{\beta}_{{\rm c}_1}+\vec{\beta}_{{\rm c}_2}+\dots \neq 0\;,
\end{equation}
while configurations with $\vec{\beta}_{{\rm c}_1}+\vec{\beta}_{{\rm c}_2}+\dots = 0$ are exclusively formed by center vortices. 
For example, in $SU(2)$, a single center vortex (Fig. 1) and a pair of center vortices attached to a monopole (Fig. 2), that correspond to zero and two turns of the off-diagonal components when we go around $\partial D_1 \circ \partial D_2$, cannot be continuously deformed one into the other.

\subsection{Charges of junctions}
\label{cj}

For junctions containing a mon\-o\-pole, it is natural defining a monopole asymptotic region $R^3-S^3$, the complement of a ball, where the quotient does not contribute to the energy, that is, the fields are stabilized so they satisfy,
\begin{equation}
\psi_q(x)=v_c\, n_q
\makebox[.3in]{,}
D_i n_q = 0
\makebox[.3in]{,}
n_q=ST_q S^{-1}
\;.
\label{asy-coset}
\end{equation} 

The most general $A_i $ can be locally written as ${\cal A}^S_i$, which in the asymptotic monopole region must satisfy
$D_i({\cal A}) T_q = 0$, that is,
\begin{equation}
[{\cal A}_i, T_q]=0\;.
\end{equation}
As this is valid for every $q=1,\dots, r$, and the $T_q$'s form a maximal set of independent elements in the algebra, commuting among them, ${\cal A}_i$ must be a combination of the $T_q$'s, 
\begin{equation}
{\cal A}_i= {\cal A}^q_i\, T_q
\makebox[.5in]{,}
A_i = {\cal A}^q_i\, ST_q S^{-1} + \frac{i}{g} S \partial_i S^{-1}\;,
\label{asy-coset-A}
\end{equation}
or defining $ \bar{A}^q_i ={\cal A}^q_i -C^q_i$, and using eq. (\ref{ansatz}), this implies,
\begin{equation}
A_i \to \bar{A}^q_i n_q - C^a_i n_a \;.
\label{asymp-mon}
\end{equation}
Then, using eq. (\ref{field-st}), in $R^3-S^3$ the local diagonal and off-diagonal components of the field strength are,
\begin{eqnarray}
G^q_{i j} & =& (\partial_i {\cal A}^q_j -\partial_j {\cal A}^q_i )- F^q_{i j}(C)
\nonumber \\
& =& (\partial_i \bar{A}^q_j -\partial_j \bar{A}^q_i )- g f^{qab} C^a_i C^b_j\;, \nonumber \\
G^a_{i j}  & = & - F^a_{i j}(C)\;.
\end{eqnarray}
From eq. (\ref{FCe}), around a given vortex of the junction, located at ${\rm c}={\rm c}_1,{\rm c}_2,\dots$, we have (cf. eq. (\ref{R-disk})),
\begin{eqnarray}
F^A_{i j}(C) &=& \frac{i}{g} \, {\rm tr}\, (M^A
e^{-i\chi_{\rm c}\,\vec{\beta}_{\rm c}\cdot \vec{M}} \, {\cal R}_{\rm c}^{-1}   [\partial_i,\partial_j]({\cal R}_{\rm c }\,   e^{i\chi_{\rm c}\, \vec{\beta}_{\rm c}\cdot \vec{M}}) \nonumber \\
&=& -\frac{1}{g} \, {\rm tr}\, (M^A
  \vec{\beta}_{\rm c}\cdot \vec{M})\, [\partial_i,\partial_j] \chi_{\rm c} \;,
\label{onvor}
\end{eqnarray}
That is, in $R^3-S^3$ the field strength is,
\begin{equation}
F_{i j} = G^q_{i j}\, n_q\;,
\end{equation}
\begin{equation}
G^q_{i j}= [(\partial_i \bar{A}^q_j -\partial_j  \bar{A}^q_i )+H^q_{i j}]
\makebox[.5in]{,}
H^q_{i j}= - g f^{qab} C^a_i C^b_j \;.
\label{Fcomp}
\end{equation}

To put in evidence the constraints implied by the exact homotopy sequence (\ref{sequence}),  let us compute the charge of the monopole along the diagonal local direction $n_q$. Using eqs. (\ref{Bfield}), (\ref{Fcomp}), it is given by,
\begin{equation}
Q_m^q=\oint dS_i\,  {\cal G}_{i}^q  = \oint dS_i\, h^q_i 
\makebox[.5in]{,}
h^q_i=\frac{1}{2}\, \epsilon_{ijk} H^q_{jk}\;,
\label{hpeq}
\end{equation}
where the flux is computed through a closed surface contained in the asymptotic monopole region. On the other hand, taking into account that a part of this region is also an asymptotic region of local minima, where $F_{jk}=0$, we can also compute the total flux as the sum of the fluxes through the disks $D_{\rm c}$, chosen
so that the $\partial D_{\rm c}$'s encircle the vortex cores. That is, we can write,
\begin{eqnarray}
\lefteqn{Q_m^q =\sum  \int dS_i\,  {\cal G}_{i}^q =\sum \int dS_i\,  (\epsilon_{ijk} \partial_j \bar{A}^q_k +h^q_i)}\nonumber \\
&&= -\sum \left(\oint dx_i\, C^q_i +\int dS_i \, h^q_i \right) = -\sum \int dS_i\,  \frac{1}{2}\, \epsilon_{ijk} F^q_{jk}(C)\;. 
\label{calc-charge}
\end{eqnarray}
Now, taking into account eq. (\ref{onvor}), the flux for each vortex is given by, 
\begin{eqnarray}
\vec{Q}_v|^q&=&-\int dS_i\,  \frac{1}{2}\, \epsilon_{ijk} F^q_{jk}(C)  \nonumber \\
&=& \frac{1}{g}\, {\rm tr}\, (M^q
\vec{\beta}_{\rm c}\cdot \vec{M}) \int dS_i\,   \epsilon_{ijk}\,  \partial_j \partial_k\chi_{\rm c} =\frac{2\pi}{g}\vec{\beta}_{\rm c}|^{q}\;,
\end{eqnarray}
so the possible monopole charges, organized as $\vec{Q}_m =(Q^1_m,\dots, Q^r_m)$, become
\begin{equation}
\vec{Q}_m  = \frac{2\pi}{g}  \sum_{\rm c} \vec{\beta}_{\rm c} \;.
\label{MC}
\end{equation}
They correspond to $(2\pi/g)\, 2\bar{N}$ times a nonzero element of the dual root lattice $\Lambda(Ad(G^{\sf v}))$ (cf. eq. (\ref{possible})). Note that among the possible monopoles, those with charge given by the roots of the dual algebra $\mathfrak{g}^{\sf v}$
are expected to be elementary, as these roots generate the lattice. 

Note also that a regular gauge transformation corresponds to $R\to {\cal R} R$, where ${\cal R}$ is regular and $R$ defines the frame defects. Then, as the singular term $F^A_{i j}(C)$ is not altered by this operation (cf. eq. (\ref{FCe})), the center vortex and monopole charge, along the local diagonal directions, are in fact invariant under these transformations.

\section{SU(N)}
\label{rev-sun}

In this section, we review some basic properties of the Cartan decomposition, and the associated roots and weights of $\mathfrak{su}(N)$. We shall follow ref. \cite{Giorgi}, but with a modified convention for the generators, so as to be consistent with the normalization we choosed in eq. (\ref{alg-norm}).

\subsection{Cartan decomposition of $\mathfrak{su}(N)$}

The rank of $\mathfrak{su}(N)$ is $r=N-1$, its dimension is $d=N^2-1$, and there are $N(N-1)/2$ pairs of root vectors $E_{\alpha}, E_{-\alpha}$. In the fundamental representation, the diagonal of $T_q$ can be given by,
\begin{equation}
\frac{1}{\sqrt{2q(q+1)N}}\,(1,\dots,1,-q,0,\dots,0)\;,
\end{equation}
where the initial $q$ elements are equal to $1$. In this case, it is convenient to define a weight as positive if the last nonvanishing component is positive, so that the weights of the fundamental matrix representation, $\vec{w}_{q}$, $q=1,\dots,N$, given by,
\begin{equation} 
\vec{\omega}_{q} = (T_1|_{qq}, T_2|_{qq},\dots, T_{N-1}|_{qq})
\makebox[.5in]{,}
\vec{w}_1+\dots +\vec{w}_N=0
\;,
\label{wfun}
\end{equation}
become ordered according to $\vec{w}_1 > \vec{w}_2 > \dots > \vec{w}_N$ , and satisfy,
\begin{equation}
\vec{w}_q\cdot \vec{\omega}_q = \frac{N-1}{2N^2}
\makebox[.5in]{,}
\vec{w}_q \cdot \vec{w}_p = -\frac{1}{2N^2}
\makebox[.3in]{,}
q\neq p\;.
\label{fundwei}
\end{equation}
The positive roots are,  
\begin{equation}
\vec{\alpha}_{qp}=\vec{w}_q -\vec{w}_p\;,
\label{difw}
\end{equation}
with $q<p$; they satisfy $\alpha^2=\vec{\alpha} \cdot \vec{\alpha}=1/N$. This is a set of $N(N-1)/2$ elements that can be ordered in the form $\vec{\alpha}_1>\vec{\alpha}_2>\dots$, 
 using the above mentioned notion of positivity.
With our conventions, the sum over all the roots (positive and negative) is,
\begin{equation}
\sum_{\alpha} \vec{\alpha}|_q \vec{\alpha}|_p = \delta_{qp}\;.
\end{equation}

In the fundamental representation, the root vector $E_\alpha$ is an $N\times N$ matrix that, for $\vec{\alpha}=\vec{\alpha}_{qp}$, only has a nontrivial element at position $qp$, whose value is $1/\sqrt{2N}$.

For example, in $SU(2)$, the weights are one component. Those of the fundamental representation are $\omega_1=\frac{1}{2\sqrt{2}}$, $\omega_2=-\frac{1}{2\sqrt{2}}$, and there is a positive root $\alpha_1=w_1 -w_2=1/\sqrt{2}$. The relation with the usual generators $\tau^A/2$, 
where $\tau^A$ are the Pauli matrices, is as follows,
\begin{equation}
T_1=\frac{\tau_3}{2\sqrt{2}} \makebox[.5in]{,} 
T_{\alpha_1}=\frac{\tau_1}{2\sqrt{2}}
\makebox[.5in]{,} T_{\bar{\alpha}_1}=\frac{\tau_2}{2\sqrt{2}}
\;. 
\end{equation}
In $SU(3)$, the weights of the fundamental representation are,
\begin{equation}
\vec{w}_ 1=(1/2\sqrt{3},1/6) \makebox[.3in]{,}
\vec{w}_2=(-1/2\sqrt{3},1/6) \makebox[.3in]{,}
\vec{w}_3=(0,-1/3) \;.
\end{equation}
Then, the positive roots are,
\begin{equation}
\vec{\alpha}_{13}=(1/2\sqrt{3},1/2) \makebox[.3in]{,}
\vec{\alpha}_{23}=(-1/2\sqrt{3},1/2) \makebox[.3in]{,}
\vec{\alpha}_{12}=(1/\sqrt{3},0) \;,
\label{roots3}
\end{equation}
whose ordered form ($\vec{\alpha}_{1} > \vec{\alpha}_{2} > \vec{\alpha}_{3}$) is given by,
\begin{equation}
\vec{\alpha}_1 = \vec{\alpha}_{13}\makebox[.3in]{,}
\vec{\alpha}_2 = \vec{\alpha}_{23}\makebox[.3in]{,}
\vec{\alpha}_3 = \vec{\alpha}_{12}\;.
\label{alpha-def}
\end{equation}
The relation with the generators $\lambda^A/2$, where $\lambda^A$ are the Gell-mann matrices, is given by,
\begin{equation}
T_1=\frac{\lambda_3}{2\sqrt{3}} \makebox[.3in]{,} T_2=\frac{\lambda_8}{2\sqrt{3}}\;,
\end{equation}
\[
T_{\alpha_1}=\frac{\lambda_4}{2\sqrt{3}}\makebox[.5in]{,}
T_{\alpha_2}=\frac{\lambda_6}{2\sqrt{3}}\makebox[.5in]{,}
T_{\alpha_3}=\frac{\lambda_1}{2\sqrt{3}}\;,\nonumber
\]
\begin{equation}
T_{\bar{\alpha}_1}=\frac{\lambda_5}{2\sqrt{3}}
\makebox[.5in]{,}
T_{\bar{\alpha}_2}=\frac{\lambda_7}{2\sqrt{3}}
\makebox[.5in]{,}
T_{\bar{\alpha}_3}=\frac{\lambda_2}{2\sqrt{3}}\;.
\end{equation}
Finally, a part of the $SU(3)$ Lie algebra can be written as,
\begin{equation}
[T_q,T_\alpha] = i\vec{\alpha}|_q T_{\bar{\alpha}}
\makebox[.3in]{,}
[T_q,T_{\bar{\alpha}}] = -i\vec{\alpha}|_q T_{\alpha}
\makebox[.3in]{,}
[T_\alpha,T_{\bar{\alpha}}] = i\vec{\alpha}|_q T_q\;,
\end{equation}
corresponding to a structure constant $f_{\alpha \bar{\alpha}q}=\vec{\alpha}|_q$. 
The indices $\alpha$ and $\bar{\alpha}$ refer to the same 
root $\vec{\alpha}$, but associated with the generators $T_\alpha$ and $T_{\bar{\alpha}}$, respectively.
The remaining commutators, with $\alpha \neq {\gamma}$ are,
\begin{equation}
 [T_\alpha,T_{\gamma}]= i f_{\alpha \gamma \bar{\delta}} T_{\bar{\delta}}
\makebox[.3in]{,}
 [T_{\bar{\alpha}},T_{\bar{\gamma}}]= i f_{\bar{\alpha} \bar{\gamma} \bar{\delta}} T_{\bar{\delta}}
\makebox[.3in]{,}
 [T_\alpha,T_{\bar{\gamma}}]= i f_{\alpha \bar{\gamma} \delta} T_{\delta}\;,
\end{equation}
with the nontrivial structure constants given by,
\begin{equation}
f_{\alpha_1 \alpha_2 \bar{\alpha}_3}= f_{\alpha_1 \bar{\alpha}_2 \alpha_3} = f_{\bar{\alpha}_1 \bar{\alpha}_2 \bar{\alpha}_3} =f_{\alpha_3 \alpha_2 \bar{\alpha}_1}=
 \frac{1}{2\sqrt{3}}\;,
\end{equation}
and cyclic permutations. Here, the indices $\alpha_1, \alpha_2$ and  $\alpha_3$ refer to the roots given in eq. (\ref{roots3}).

\subsection{Weights of $Z(N)$ vortices}

For $\mathfrak{su}(N)$, $\bar{N}$ is given by $(d-r)/r=N$, and $\alpha^2=
N$, so that $\vec{\alpha}^{\,\sf v}=\alpha$, and the possible $\vec{\beta}$'s are (see \cite{GNO}, \cite{notes} and references therein),
\begin{equation}
\vec{\beta}=2N\vec{w}\;,
\label{betasw}
\end{equation}
where $\vec{w}$ are the weights of any matrix representation of $SU(N)$. 
In particular, recalling that the weights of the fundamental matrix representation satisfy eq. (\ref{fundwei}), and the roots are given by 
their differences (cf. eq. (\ref{difw})), the desired single-valuedness condition in eq. (\ref{sing-v}) can be verified for
\begin{equation}
\vec{\beta}_q=2N\vec{w}_q \;,
\label{mmu}
\end{equation}
\begin{equation}
\vec{\beta}_q\cdot \vec{\alpha}_{qp} =1
\makebox[.7in]{,}
\vec{\beta}_q\cdot \vec{\alpha}_{pp'} = 0~~(p,p' \neq q) \;.
\label{rel} 
\end{equation}

As noted in ref. \cite{Konishi-Spanu}, fundamental and antifundamental weights correspond to the minimum charges. Applying $S_v(2\pi)$ (cf. eq. (\ref{RSvort})) on any weight vector of the fundamental representation, it is obtained,
\begin{equation}
 e^ {i2\pi\, \vec{\beta}_q \cdot \vec{w}_p}= e^{i2\pi z/N}\;,
\end{equation}
and as $2\pi\, \vec{\beta}_q \cdot \vec{w}_p$ corresponds to either $- \frac{2\pi}{N}$ (if $q\neq p$) or $\frac{2\pi(N-1)}{N}\equiv - \frac{2\pi}{N}$ (if $q=p$), a topological charge $z=-1$ is implied.
The generators of the antifundamental representation $-T_A^\ast$ have associated weights $-\vec{w}_q$, thus leading to a topological charge $z=1$. 

\section{Center vortex ansatz}
\label{cva}

For a straight center vortex with fundamental weight $\vec{\beta}_q$, let us consider the extended map (cf. eq. (\ref{RSvort}))
\begin{equation}
S= e^{i\varphi\, \vec{\beta}_q \cdot \vec{T}}\;.
\end{equation}
From eq. (\ref{algebrag}), we have, 
\begin{equation}
[\vec{\beta}_q \cdot \vec{T}, E_\alpha]= (\vec{\alpha}\cdot \vec{\beta}_q)\, E_\alpha
\makebox[.5in]{,}
S E_\alpha S^{-1} = e^{i(\vec{\alpha}\cdot \vec{\beta}_q)\varphi} E_\alpha \;.
\end{equation}
Therefore, for the extended local frame, besides the trivial diagonal components,
\begin{equation}
n_q(x)=ST_q S^{-1}= T_q \;,
\end{equation}
we get 
\begin{eqnarray}
n_{\alpha}(x)&=& S T_{\alpha} S^{-1} = \cos (\vec{\alpha}\cdot \vec{\beta}_q) \varphi \, T_{\alpha} -
\sin (\vec{\alpha}\cdot \vec{\beta}_q) \varphi \, T_{\bar{\alpha}} \;,\nonumber \\
n_{\bar{\alpha}}(x)&=&S T_{\bar{\alpha}} S^{-1}= \sin (\vec{\alpha}\cdot \vec{\beta}_q) \varphi \, T_{\alpha} + 
\cos (\vec{\alpha}\cdot \vec{\beta}_q) \varphi \, T_{\bar{\alpha}} \;,
\label{rotation}
\end{eqnarray}
and using eq. (\ref{rel}), for the positive roots of the form $\vec{\alpha}=\vec{\alpha}_{pp'},~p,p' \neq q$, we have
$n_{\alpha}(x)\equiv T_{\alpha}$, $n_{\bar{\alpha}}(x)\equiv T_{\bar{\alpha}}$. On the other hand, for
those roots where the weight $\vec{w}_q$ participates, the pairs 
$n_{\alpha}(x)$, $n_{\bar{\alpha}}(x)$ rotate once when we go around the center vortex, so these frame components have a defect at the $z$-axis, where the guiding centers of the vortex are located. The cases $\vec{\alpha}=\vec{\alpha}_{qp'}$ ($q<p'$) and $\vec{\alpha}=\vec{\alpha}_{pq}$ ($p<q$) have opposite senses of rotation. Then, we shall propose the following ansatz,
\begin{equation}
\psi_q =h_{qp} T_{p}
\makebox[.5in]{,}
\psi_{\alpha}=h_{(\alpha)} S T_{(\alpha)}S^{-1}
\makebox[.5in]{,}
\psi_{\bar{\alpha}}=h_{(\bar{\alpha})} S T_{(\bar{\alpha})}S^{-1}
\;, 
\label{vortex-prof}
\end{equation}
\begin{equation}
A_i = \frac{1}{g} a(\rho) \partial_i \varphi\, \vec{\beta}_q \cdot \vec{T} \;.
\end{equation}
which corresponds to,
\begin{equation}
{\cal A}_i = \frac{1}{g}(a-1)\partial_i \varphi\, \vec{\beta}_q \cdot \vec{T}\;.
\end{equation}
From eq. (\ref{rotation}), it is natural to  assume $h_{\alpha}=h_{\bar{\alpha}}$, and use for $h_\alpha$ only two profile functions $h(\rho)$, $h_0(\rho)$, depending on whether the components $n_\alpha, n_{\bar{\alpha}}$ rotate or not. When $\rho \to \infty$, in order to have a configuration of local minima,  we require,
\begin{equation}
h_{qp}\to v_c\, \delta_{qp}
\makebox[.5in]{}
h(\rho) \to v_c
\makebox[.5in]{,}
h_0(\rho) \to v_c
\makebox[.5in]{,}
a(\rho) \to 1\;,
\end{equation}
while the regularity conditions when $\rho \to 0$ are, 
\begin{equation}
h(\rho) \to 0
\makebox[.5in]{,}
a(\rho) \to 0\;.
\end{equation}

Using this ansatz in eqs. (\ref{YM1}), (\ref{YM2}), and after putting $h_{\alpha}=h_{\bar{\alpha}}$, the equations for the gauge field and diagonal profiles become,
\begin{equation}
 \frac{1}{g}\left(\frac{a'}{\rho}- a''\right)  \vec{\beta}_q \cdot \vec{T}=
2g\, (\vec{\alpha}\cdot \vec{\beta}_q) (1-a) h^2_{\alpha}\, \vec{\alpha}\cdot \vec{T} 
\;,
\label{Max}
\end{equation}
\begin{eqnarray}
(\partial^2 h_{qp})T_p &=&\mu^2 h_{qp} T_p + 2 (\kappa\,\vec{\alpha}|_q + \lambda\, \vec{\alpha}|_{q'}  h_{qq'} ) h^2_\alpha\, \vec{\alpha}|_p T_p\;,
\label{diagon-sec}
\end{eqnarray}
with summation over repeated indices, $q'$, $p$, and $\vec{\alpha}$. The sum over $\vec{\alpha}$ is only done over the positive roots.
From the field equation for $\psi_\alpha$, we obtain,
\begin{eqnarray}
\lefteqn{(\partial^2 h_{(\alpha)} -(1-a)^2 (\vec{\alpha}\cdot \vec{\beta}_q)^2 (1/\rho^2)  h_{(\alpha)})T_{(\alpha)}=}\nonumber \\
&&= (\mu^2 h_{(\alpha)}  + 2\kappa\, h_{(\alpha)}h_{qq'} \vec{\alpha}|_q \vec{\alpha}|_{q'}+ \lambda\, h^3_{(\alpha)} \alpha^2) T_{(\alpha)}
\nonumber \\
&&+\lambda\, h^2_{\gamma}h_{(\alpha)} T_{\gamma} \wedge (T_{(\alpha)} \wedge T_{\gamma})+\lambda\, h^2_{\gamma}h_{(\alpha)} T_{\bar{\gamma}} \wedge (T_{(\alpha)} \wedge T_{\bar{\gamma}}) 
\nonumber \\
&&+2\kappa\, h_{\gamma} h_{\delta}\, f_{(\alpha) \gamma \bar{\delta}}  f_{\gamma \bar{\delta} (\alpha)} T_{(\alpha)} + \lambda\, \vec{\alpha}|_{q'} \vec{\alpha}|_{q''} h_{(\alpha)} h_{qq'} h_{qq''} T_{(\alpha)}\;.\nonumber \\
\end{eqnarray}
Here, the sum over $\gamma$ is done with $\gamma \neq \alpha$, and there is no summation over $\alpha$. As   $h_{\alpha}=h_{\bar{\alpha}}$, after working out the algebra, the field equation for $\psi_{\bar{\alpha}}$ gives no new information. 

\subsection{SU(2)}

In $SU(2)$ we have a pair of one-component weights. Taking $S= e^{i\varphi\, \sqrt{2} T_1}$, 
the ansatz for the Higgs fields, and the equations of motion are ($h_{11}= h_1$),
\begin{equation}
\psi_1 =h_{1}\, T_{1}\;,
\makebox[.5in]{,} 
\psi_{\alpha_1}= h\, S T_{\alpha_1} S^{-1}\;,
\makebox[.5in]{,}
\psi_{\bar{\alpha}_1}= h\, S T_{\bar{\alpha}_1} S^{-1}
\;,
\end{equation}
\begin{equation}
 \frac{1}{g}\left(\frac{a'}{\rho}- a''\right)  =
g\, (1-a) h^2\;,
\end{equation}
\begin{equation}
h''_{1}+(1/\rho)h'_{1}= \mu^2 h_{1} + (\kappa +\lambda h_{1}) h^ 2\;,
\end{equation}
\begin{equation}
h''+(1/\rho)h'-(1-a)^2(1/\rho^2) \,h = \mu^2 h + \kappa h h_{1} + \frac{\lambda}{2} h(h^ 2+h^2_{1})
\end{equation}
This is essentially a generalized set of Nielsen-Olesen equations extended by the presence of a profile function $h_1(\rho)$ that is not required to vanish at the origin, and tends to $v_c$, when $\rho \to \infty$.

\subsection{SU(3)}
For definiteness, let us consider a vortex with weight $\vec{\beta}_1=(\sqrt{3}/2,1)$. Using that, 
$\vec{\alpha}_1\cdot \vec{\beta}_1=1$, $\vec{\alpha}_2\cdot \vec{\beta}_1=0$,  and 
$\vec{\alpha}_3\cdot \vec{\beta}_1=1$, the off-diagonal fields in eq. (\ref{vortex-prof}) become,
\begin{equation}
\psi_{\alpha_1}= h\, S T_{\alpha_1} S^{-1}
\makebox[.5in]{,}
\psi_{\alpha_2}=  h_0\, T_{\alpha_{2}} 
\makebox[.5in]{,}
\psi_{\alpha_3}= h\, S T_{\alpha_3} S^{-1}
\;,
\end{equation}
\begin{equation}
\psi_{\bar{\alpha}_1}= h\, S T_{\bar{\alpha}_1} S^{-1}
\makebox[.5in]{,}
\psi_{\bar{\alpha}_2}=  h_0\, T_{\bar{\alpha}_{2}} 
\makebox[.5in]{,}
\psi_{\bar{\alpha}_3}= h\, S T_{\bar{\alpha}_3} S^{-1}
\;.
\end{equation}
The use of the same profile function for $\psi_{\alpha_1}$ and $\psi_{\alpha_3}$ is necessary in order for the left and right hand sides in eq. (\ref{Max}) be oriented along the same direction in the algebra. In this case, the sum over $\vec{\alpha}$ gives,
\begin{eqnarray}
2g\, (\vec{\alpha}\cdot \vec{\beta}_q) (1-a) h^2_{\alpha}\, \vec{\alpha}\cdot \vec{T} &=&2g\, (1-a) h^2\, \, (\vec{\alpha}_1+\vec{\alpha}_3)\cdot \vec{T}\nonumber \\
&=&
g\,  (1-a)h^2\, \vec{\beta}_1\cdot \vec{T}
\;,
\end{eqnarray}
and replacing in eq. (\ref{Max}), we get,
\begin{equation}
 \frac{1}{g}\left(\frac{a'}{\rho}- a''\right) = 
g\,  (1-a) h^2
\;.
\label{su3eq}
\end{equation}

Now, when summing over positive roots, we have $\vec{\alpha}|_q \vec{\alpha}|_p = \frac{1}{2}\delta_{qp}$, and the field equation (\ref{diagon-sec}), associated with Higgs fields along diagonal directions, turns out to be,
\begin{eqnarray}
\partial^2 h_{qp} &=&\mu^2 h_{qp} +  \kappa h^2\, \delta_{qp}+ \lambda h^2\,  h_{qp}  \nonumber \\
&&+ 2 (\kappa\,\vec{\alpha}_2|_q + \lambda\, \vec{\alpha}_2|_{q'}  h_{qq'} ) (h_0^2-h^2)\, \vec{\alpha}_2|_p\;.
\end{eqnarray}
The equation for $\psi_\alpha$ leads to,
\begin{eqnarray}
&&\partial^2 h_{\alpha} -(1-a)^2 (\vec{\alpha}\cdot \vec{\beta}_1)^2 (1/\rho^2)  h_{\alpha}=\nonumber \\
&=& \mu^2 h_\alpha  + 2\kappa\, h_{\alpha}h_{qq'} \vec{\alpha}|_q \vec{\alpha}|_{q'}+ (\lambda/3)\, h^3_{\alpha}
+(\lambda/6)\, h_{\alpha}\sum_{\gamma \neq \alpha} h^2_{\gamma}  \nonumber \\
&&+2\kappa\, h_{\gamma} h_{\delta}\, f_{(\alpha) \gamma \bar{\delta}}  f_{\gamma \bar{\delta} (\alpha)} + \lambda\, \vec{\alpha}|_{q'} \vec{\alpha}|_{q''} h_{(\alpha)} h_{qq'} h_{qq''} \;
\end{eqnarray}
(with summation over $q$, $q'$, and no sum over $\vec{\alpha}$). The $\vec{\alpha}_1$ component implies,
\begin{eqnarray}
&&\partial^2 h -(1-a)^2  (1/\rho^2)  h=\nonumber \\
&=& \mu^2 h  + 2\kappa\, h\, h_{qq'} \vec{\alpha}_1|_q \vec{\alpha}_1|_{q'}+ (\lambda/3)\, h^3
+(\lambda/6)\, h (h^2+h_0^2)  \nonumber \\
&&+(\kappa/3)\, h\,h_0   + \lambda\, \vec{\alpha}_1|_{q'} \vec{\alpha}_1|_{q''} h\, h_{qq'} h_{qq''} \;.
\end{eqnarray}
while for the $\vec{\alpha}_2$ component, we obtain,
\begin{eqnarray}
\partial^2 h_0 
&=& \mu^2 h_0  + 2\kappa\, h_0 h_{qq'} \vec{\alpha}_2|_q \vec{\alpha}_2|_{q'}+ (\lambda/3)\, h^3_0
+(\lambda/3)\, h_0 h^2 \nonumber \\
&&+(\kappa/3) h^2+ \lambda\, \vec{\alpha}_2|_{q'} \vec{\alpha}_2|_{q''} h_0 h_{qq'} h_{qq''} \;.
\end{eqnarray}
The equation for $\vec{\alpha}_3$ is similar to that for $\vec{\alpha}_1$, with $\vec{\alpha}_3$ in the place of $\vec{\alpha}_1$.

Now, considering the following structure,
\begin{equation}
h_{qp}= \frac{1}{4} h_1\, \vec{\beta}_1|_q \vec{\beta}_1|_p + 3 h_2\, \vec{\alpha}_2|_q \vec{\alpha}_2|_p\;,
\end{equation}
and using,
\begin{equation}
\delta_{qp}= \frac{1}{4}\, \vec{\beta}_1|_q \vec{\beta}_1|_p + 3 \, \vec{\alpha}_2|_q \vec{\alpha}_2|_p\;,
\label{delta-comp}
\end{equation}
the equations for the Higgs profiles accompanying frame components that have no defects become simplified to the form,
\begin{eqnarray}
h''_0+(1/\rho)h'_0 
&=& \mu^2 h_0  + (\kappa/3) \,( 2h_0 h_2 +h^2) + (\lambda/3)\,h_0  (h^2_0 + h^2+  h_2^2)  \;,\nonumber \\
h''_1+(1/\rho)h'_1 &=&  \mu^2 h_1 +    \kappa h^2 + \lambda  h^2 \,h_1 \;, \nonumber \\
h''_2+(1/\rho)h'_2 &=&  \mu^2 h_2 +  \kappa h^2 + \lambda h^2\, h_2 +(2/3)(\kappa + \lambda h_2)\,  (h_0^2-h^2)\;.
\end{eqnarray}
From eq. (\ref{delta-comp}), the boundary conditions for $h_{qp}$ amount to $h_1\to v_c$, $h_2\to v_c$, when $\rho \to \infty$. 

Finally, the equations for $\vec{\alpha}_1$ and $\vec{\alpha}_3$, that correspond to the frame components having defects, become simplified to a single equation for the profile function $h$,
\begin{eqnarray}
\lefteqn{h''+(1/\rho)h' -(1-a)^2  (1/\rho^2)  h = \mu^2 h } \nonumber \\
&&+(\kappa/6) h\, (2 h_0 +  3h_1 +  h_2 )+ (\lambda/12) h \, (  6\, h^2 +2h_0^2+  3h_1^2+  h_2^2 )\;.
\end{eqnarray}
Therefore, we obtain again a modified set of Nielsen-Olesen equations, but coupled to three profile functions $h_0, h_1, h_2$, which are not required to vanish at the origin.

\section{Junctions}
\label{complex}

In this section, we shall write the non Abelian phase factors corresponding to different types of junctions. They will provide the guiding 
centers, and the charges, or fluxes, of center vortices and monopoles. These fluxes will be defined with respect to the outward normal to the 
surface $S^2$ around the center of the junction. These factors will also define those gauge and Higgs field components that have defects. Therefore, they are fundamental to obtain the appropriate boundary conditions, as well as determining the type of quark sources they can bind. 

Before doing this, we would like to comment about the smoothness of the gauge field behavior in the asymptotic monopole region, given in eq. (\ref{asymp-mon}), and the form of the monopole density current. 
As the off-diagonal fields $C^a_i$ depend on frame components containing defects, it is important to check whether or not these singularities are cancelled in the second term, $C^a_i n_a$.
From eq. (\ref{CAm}), and using 
$$R^{-1}\partial_i R|_{BC} =\hat{e}_B \cdot R^{-1}\partial_i R\, \hat{e}_C=  \hat{n}_B \cdot \partial_i \hat{n}_C\;,$$ we have,
\begin{equation}
C^A_i\, M_A|_{BC} =\frac{i}{g} R^{-1} \partial_i  R|_{BC}
\makebox[.5in]{,}
-g\, f_{ABC}\, C_i^A=  \langle n_B , \partial_i
n_C\rangle \;.
\label{Cprop}
\end{equation}
Then, we can write,
\begin{equation}
n_q \wedge \partial_i n_p =  \langle n_A, \partial_i n_p \rangle\, n_q \wedge n_A 
= - g\, f_{DAp}\,C_i^D\,  f_{qAB}\, n_B \;,
\end{equation}
and defining, 
\begin{equation}
P_\beta =\vec{\beta}|^q n_q = S\, \vec{\beta}\cdot \vec{T} S^{-1}\;,
\label{btriv}
\end{equation}
where the $\vec{\beta}$'s are ($2N$ times) weights of an irreducible representation, we obtain,
\begin{eqnarray}
P_\beta \wedge \partial_i P_\beta &=& g\, (\vec{\beta}\cdot \vec{M})^2|_{ab} \,C_i^a\,  n_b \;.
\end{eqnarray}

Now, for $SU(N)$ we have the property,
\begin{equation}
\sum_{\beta}(\vec{\beta}\cdot \vec{M})^2|_{ab} = c_r \, \delta_{ab}\;,
\label{suma}
\end{equation}
where $c_r$ only depends on the irreducible representation of $SU(N)$ considered; for the fundamental representation, $c_r=2$. This can be verified by acting with both members on the nonzero root vectors, and using the Freudenthal - de Vries formula for $\sum_{\vec{w}} (\vec{w}\cdot \vec{\alpha})^2 $ \cite{Burns, Humphreys}. Thus, we get,
\begin{equation}
C_i^a\,  n_a 
= (1/gc_r) \sum_{\beta} P_\beta \wedge \partial_i P_\beta  \;,
\label{Cnw}
\end{equation}
that is, the gauge field in the asymptotic monopole region (cf. eq. (\ref{asymp-mon})) can be written in terms of $\bar{A}^q_i$ and the diagonal components $n_q$, which are regular on that region. Note also that the set of weights is invariant under the Weyl group \cite{GNO,Brian}.

We can proceed in a similar manner to get information about the monopole density current $h^q_i$ 
in eq. (\ref{hpeq}), that is given in terms of $H^q_{ij}$ in eq. (\ref{Fcomp}). From the general property,
\begin{eqnarray}
\langle n_C , \partial_i n_A \wedge \partial_j n_B\rangle &=&
\langle n_C , n_{A'}\wedge n_{B'} \rangle \langle n_{A'}, \partial_i n_A\rangle \langle n_{B'}, \partial_j n_B\rangle \nonumber \\
&=& ig^2 M_A M_C M_B|_{DE} \,C^D_\mu  C^E_\nu \;,
\end{eqnarray}
where we used the algebra and eq. (\ref{Cprop}), we obtain,
\begin{equation}
\langle P_{\beta}, \partial_i P_{\beta} \wedge \partial_j P_{\beta} \rangle
= i g^2  (\vec{\beta} \cdot  \vec{M})^3|_{DE} \,C^D_\mu  C^E_\nu \;.
\end{equation}
For a fundamental weight $\vec{\beta}$, we have, $(\vec{\beta}\cdot  \vec{M})^3 =  \vec{\beta}\cdot  \vec{M}$, as both members give the same value $(\vec{\beta}\cdot \vec{\alpha})^3=\vec{\beta}\cdot \vec{\alpha}$ when applied on the nonzero root vectors (cf. eq. (\ref{rel})), and vanish when applied on the zero root vectors. Therefore, we get,
\begin{equation}
\vec{\beta}|^q H^q_{ij}= -g \vec{\beta}|^q f_{qde} \,C^d_\mu  C^e_\nu =  -\frac{1}{g}\langle P_{\beta}, \partial_i P_{\beta} \wedge \partial_j P_{\beta} \rangle
 \;.
\label{topo-curr}
\end{equation}
That is, the projections of the densities $H^q_{ij}$ only depend on the coset, they are smooth everywhere but at the pointlike defects in the local diagonal sector, and have the same form as the well-known topological charge densities for $S^2 \to S^2$ mappings. 

\subsection{$N$-branch normal junction}

The normal junction 
is simply formed by assembling center vortices or confining strings. In \S \ref{appa}, we shall see the natural way 
to introduce quark probes in the dual model. After this, it will become clear that this type of junction confines a color singlet state of $N$ external fundamental quarks to form a normal baryon. The normal junction can be defined by the mapping,
\begin{equation}
S = e^{i(\varphi_1\, \vec{\beta}_1+\dots + \varphi_{N-1}\, \vec{\beta}_{N-1}) \cdot \vec{T}}\;,
\label{normalY}
\end{equation}
where, for $q=1,\dots,N-1$, the weights $\vec{\beta}_q$ are given by eq. (\ref{mmu}), and the multivalued phases $\varphi_q$  satisfy,
\begin{equation}
J_q|_i =\epsilon_{ijk} \partial_j \partial_k \varphi_q
\makebox[.5in]{,}
J_q = \int ds\,  \frac{dx_q}{ds}\, \delta^{(3)}(x- x_q(s))\;,
\end{equation}
$x_q\in R^3$.
For each path, the parameter $s$ runs from $-\infty$ to $+\infty$; in these limits the paths tend to infinity in $R^3$, and they are chosen as follows. In the interval $s\in (-\infty,0]$ all the paths coincide, describing the $N$-th  branch, running from infinity to the point O at $s=0$ (the center of the junction). 
A surface $S^2$ around O is intersected by this branch at a guiding center ${\rm c}_N$. Then, for parameters in the interval $s \in [0, +\infty )$, the $N-1$ paths separate forming the other $N-1$ branches that intersect $S^2$ at the guiding centers ${\rm c}_q$, $q=1,\dots N-1$. 

The mapping $R=R(S)$ is ill-defined at the centers $c_1, \dots, c_N$. 
On the disks $D_{{\rm c}_q}$ centered at ${\rm c}_q$, $q=1,\dots,N-1$, $S(x)$ behaves as,
\begin{equation}
S_{{\rm c}_q}(x) \sim  {\cal S}_{{\rm c}_q} \,  e^{i\varphi_{(q)}\, \vec{\beta}_{(q)} \cdot \vec{T}}
\makebox[.5in]{,}
{\cal S}_{{\rm c}_q} = \prod_{p,\, p\neq q} e^{i\varphi_p\, \vec{\beta}_p \cdot \vec{T}}
\;
\end{equation}
where the factor ${\cal S}_{{\rm c}_q}$ is smooth, as it depends on phases $\varphi_p$, $p\neq q$, which are regular at ${\rm c}_q$.
On the disk $D_{{\rm c}_N}$ centered at ${\rm c}_N$, when its radius $\epsilon$ gets smaller, all the phases have approximately the same value, $\varphi_q \sim \varphi $, $q=1,\dots, N-1$, thus implying the following behavior,
\begin{equation}
S_{{\rm c}_N}(x) \sim  e^{i\varphi \, (\vec{\beta}_1 + \dots + \vec{\beta}_{N-1}) \cdot \vec{T}} = 
e^{-i\varphi \, \vec{\beta}_N \cdot \vec{T}}\;.
\end{equation}
Then, comparing with eq. (\ref{R-disk}), we can identify,
\begin{equation}
\chi_{{\rm c}_q} \sim \varphi_q
\makebox[.5in]{,}
\vec{\beta}_{{\rm c}_q} = \vec{\beta}_{q}
\makebox[.5in]{,}
q=1,\dots,N-1\;,
\end{equation}
\begin{equation}
\chi_{{\rm c}_N} \sim -\varphi
\makebox[.5in]{,}
\vec{\beta}_{{\rm c}_N} = \vec{\beta}_N\;.
\end{equation}
The center vortex charges, labeled by $\vec{\beta}_{{\rm c}_1}, \dots, \vec{\beta}_{{\rm c}_N}$, then correspond to the $N$ weights of the fundamental representation, satisfying the condition $\vec{\beta}_{{\rm c}_1}+\dots +\vec{\beta}_{{\rm c}_N}=0$. This means that there is no monopole at the origin, and the junction will match $N$ fundamental quarks having the $N$ charges different.

It is also worth mentioning that the same mapping $S$ can be rewritten in terms of any set of 
$N-1$ independent fundamental weights, by simply changing variables to a set of phases whose associated currents coincide on a different branch.

\subsection{Two-branch vortex-monopole junction}
\label{two-branch}

This junction is formed by a pair of center vortices attached to a monopole, one of the vortex charges belongs to the fundamental representation, while the other belongs to the antifundamental. When adding external quark sources in the dual YMH model, 
this junction will be able to confine a color nonsinglet quark/antiquark pair that becomes bound to a valence gluon to form a hybrid meson state. 
For general $SU(N)$, it can be defined by the mapping,
\begin{equation}
S= e^{i\varphi\, \vec{\beta}_q \cdot \vec{T}}\, W(x)
\makebox[.5in]{,}
W(x)=e^{i\theta \,\sqrt{N} T_{\alpha_{qp}}} \;,
\label{ele-comp}
\end{equation}
where $\varphi$ and $\theta$ are the polar angles, and the roots and fundamental weights are given by eqs. (\ref{difw}), (\ref{mmu}). 
At the north pole, the behavior is, 
\begin{equation}
S_{{\rm c}_1}(x) \sim  e^{i\chi_{{\rm c}_1}\, \vec{\beta}_q \cdot \vec{T}}
\makebox[.5in]{,}
\chi_{{\rm c}_1} \sim \varphi\;.
\label{n-beh}
\end{equation}
At the south pole, $W(x)$ approaches $W_{\alpha}=e^{i\pi \,\sqrt{N} T_{\alpha}}$, a Weyl reflection whose effect is,
\begin{equation}
W^{-1}_{\alpha} \vec{\beta}\cdot \vec{T}\, W_{\alpha}=\vec{\beta}' \cdot \vec{T} \;. 
\makebox[.5in]{,}
\vec{\beta}'= \vec{\beta} - 2 \vec{\alpha}\, (\vec{\alpha}\cdot \vec{\beta})/ \alpha^2\;.
\end{equation}
In our case, where $\vec{\beta}=\vec{\beta}_q$ and $\vec{\alpha}=\vec{\alpha}_{qp}$,
\begin{eqnarray}
\vec{\beta}'&=& \vec{\beta}_q - 2N \vec{\alpha}_{qp}\, (\vec{\alpha}_{qp}\cdot \vec{\beta}_q)\nonumber \\
&=& \vec{\beta}_q - 2N (\vec{w}_{q} -\vec{w}_{p})=\,\,\,\vec{\beta}_p\;,
\end{eqnarray}
so that the behavior at the south pole is,
\begin{equation}
S_{{\rm c}_2}(x) \sim W_{\alpha}\, e^{i\varphi\, \vec{\beta}_p \cdot \vec{T}}= W_{\alpha}\, e^{-i\chi_{{\rm c}_2}\, \vec{\beta}_p \cdot \vec{T}} \makebox[.5in]{,}
\chi_{{\rm c}_2} \sim -\varphi
\;.
\label{s-beh}
\end{equation}
Then, comparing (\ref{n-beh}) and (\ref{s-beh}) with eq. (\ref{R-disk}), we obtain,
\begin{equation}
\vec{\beta}_{{\rm c}_1} = \vec{\beta}_{q}
\makebox[.5in]{,}
\vec{\beta}_{{\rm c}_2} = -\vec{\beta}_{p}\;,
\end{equation}
and the monopole charge is (cf. eq. (\ref{MC})), 
\begin{equation}
\vec{Q}_m= \frac{2\pi}{g}\, (\vec{\beta}_{{\rm c}_1}+\vec{\beta}_{{\rm c}_2})=\frac{2\pi}{g}\, 2N (\vec{w}_q-\vec{w}_p) = \frac{2\pi}{g}\, 2N \vec{\alpha}_{qp}\;.
\end{equation}
As the roots $\vec{\alpha}_{qp}$ generate the lattice $\Lambda(Ad(SU(N)))$, the monopole in this juncion is expected to be an elementary object of the dual model. In fact, it can be naturally identified with a confined valence gluon with color $\vec{\alpha}_{qp}$. 

In the local basis,
\begin{equation}
n_q = S T_q S^{-1}
\makebox[.5in]{,}
n_{\alpha} = S T_{\alpha} S^{-1}
\makebox[.5in]{,}
n_{\bar{\alpha}} = S T_{\bar{\alpha}} S^{-1}\;,
\end{equation}
the components having a vortex-like defect at $\theta=0$ ($\theta=\pi$) are those $n_\alpha$, $n_{\bar{\alpha}}$ such that $T_\alpha$
does not commute with $\vec{\beta}_q\cdot \vec{T}$ ($\vec{\beta}_p\cdot \vec{T}$). This is the case when one of the fundamental weights forming $\vec{\alpha}$ (cf. eq. (\ref{difw})) coincides with $\vec{w}_q$ ($\vec{w}_p$). 

In the case of $SU(2)$, we can take for example,
\begin{equation}
S= e^{i\varphi\, \sqrt{2} T_1}e^{i\theta \,\sqrt{2} T_{\alpha_1}}\;,
\end{equation}
which leads to,
\begin{eqnarray}
n_1 &=& \cos \theta\, T_1 + \sin \theta \left( \cos \varphi\, T_{\bar{\alpha}_1} + \sin \varphi\, T_{\alpha_1}\right)  \nonumber \\
n_{\bar{\alpha}_1} &=& -\sin \theta \, T_1 + \cos \theta\, (\cos \varphi \,
T_{\bar{\alpha}_1}  +  \sin \varphi\,  T_{\alpha_1}) \nonumber \\
n_{\alpha_1}&=& -\sin \varphi\, T_{\bar{\alpha}_1}+ \cos \varphi\, T_{\alpha_1} 
\;,
\end{eqnarray}
and the components $n_{\alpha_1}$, $n_{\bar{\alpha}_1}$ rotate once when we approach the positive and negative $z$-axis. This is the frame depicted in Fig. 2. 

In the case of $SU(3)$, taking,
\begin{equation}
S= e^{i\varphi\, \vec{\beta}_1 \cdot \vec{T}} e^{i\theta \,\sqrt{3} T_{\alpha_3}}\;,
\end{equation}
the components $n_\alpha, n_{\bar{\alpha}}$ with $\vec{\alpha}=\vec{\alpha}_1, \vec{\alpha}_3$ 
(resp. $\vec{\alpha}_2, \vec{\alpha}_3$) contain a vortex-like defect on the positive (resp. negative) $z$-axis. For the definition of $\vec{\alpha}_1, \vec{\alpha}_2, \vec{\alpha}_3$, see eq. (\ref{alpha-def}).

For general $SU(N)$, the local diagonal directions can be analyzed by considering the projections $P_\beta$ in eq. (\ref{btriv}).
Among them, those with $\vec{\beta} \neq \vec{\beta}_q , \vec{\beta}_p$ are trivial. In addition, because of eq. (\ref{wfun}), $\vec{\beta}_q + \vec{\beta}_p$ can be written in terms of weights other than $\vec{\beta}_q$ and $\vec{\beta}_p$,  so that the combination $P_{{\beta}_q} +P_{{\beta}_p}$ is also trivial. Then, in the local Cartan sector, the nontrivial frame dependence is contained in the combination,
\begin{equation}
P_{{\beta}_q}-P_{{\beta}_p} = S (\vec{\beta}_q-\vec{\beta}_p)\cdot \vec{T}S^{-1}= 2 L_{\alpha}
\makebox[.5in]{,}
\vec{\alpha}=\vec{\alpha}_{qp}\;,
\end{equation}
\begin{equation}
L_{\alpha}=S\, \frac{\vec{\alpha} \cdot \vec{T}}{\alpha^2} S^{-1}\;.
\end{equation}
Note that the Lie algebra element $L_{\alpha}$, is a local version of the diagonal generator of the $\mathfrak{so}(3)$ subalgebra in eq. (\ref{subalg}). Using eq. (\ref{ele-comp}), we obtain,
\begin{eqnarray}
L_{\alpha} &=&  \cos \theta\, \frac{\vec{\alpha}\cdot \vec{T}}{\alpha^2} + \sin \theta \left( \cos \varphi 
\, \frac{T_{\bar{\alpha}}}{\sqrt{\alpha^2}}+ \sin \varphi 
\, \frac{T_{\alpha}}{\sqrt{\alpha^2}}\right)\;.
\label{elea}
\end{eqnarray}

In other words, the frame defined by eq. (\ref{ele-comp}) describes a center vortex with weight $\vec{\beta}_p$ entering a surface $S^2$ around the origin, a center vortex with weight $\vec{\beta}_q$ leaving, interpolated by a monopole configuration. The latter correlates $S^2$, the space of directions around the origin, $$\hat{r}=(\cos \theta , \sin \theta \cos \varphi  , \sin \theta \sin \varphi )\;,$$ with $L_{\alpha}$, a generator of a local $\mathfrak{so}(3)$ algebra labeled by $\vec{\alpha}=\vec{\alpha}_{qp}$,
\begin{equation}
L_{\alpha}\makebox[.5in]{,}
\frac{1}{\sqrt{\alpha^2}}\, n_{\alpha}
\makebox[.5in]{,}
\frac{1}{\sqrt{\alpha^2}}\, n_{\bar{\alpha}}\;.
\end{equation}
Note also that at $\theta=0$ and $\theta=\pi$, the ill-defined phase $\varphi$ is not manifested in the local Cartan sector, as required by the general discussion given in \S \ref{chains-comp}. In that sector, the only defect is at the origin, where $L_{\alpha}$ is ill-defined, so that the profile functions accompanying $L_{\alpha}$ will be required to be zero at that point.

The frame dependent fields $C^A_\mu$ in eq. (\ref{rep2}) are important for the general parametrization of the gauge field $A_\mu$ in eq. (\ref{ansatz}). Using eq. (\ref{CAm}), we get,
\begin{eqnarray}
C^A_i\, M_A &=& -\frac{1}{2g}\, \partial_i \varphi \left[(\vec{\beta}_q +\vec{\beta}_p) +\cos \theta\, (\vec{\beta}_q -\vec{\beta}_p)\right]\cdot \vec{M} \nonumber \\
&& +\frac{\sqrt{N}}{g}\, \left(\partial_i \varphi \sin \theta\, M_{\bar{\alpha}} - \partial_i \theta \, M_{\alpha}\right)\;.
\end{eqnarray}
As expected, when $\theta \to 0$ (resp. $\theta \to \pi$), the diagonal part is oriented along $\vec{\beta}_q$ (resp. $\vec{\beta}_p$). In addition, from eqs. (\ref{btriv}), (\ref{Cnw}), we have,
\begin{equation}
C_i^a\,  n_a = (1/2g) [ P_{\beta_q} \wedge \partial_i P_{\beta_q} + P_{\beta_p} \wedge \partial_i P_{\beta_p} ]=(1/g)\, L_{\alpha} \wedge \partial_i L_{\alpha} \;.
\end{equation}
Therefore, we get,
\begin{equation}
A_i =  ({\cal A}^A_i -C^A_i)\, n_A \;,
\end{equation}
\begin{equation}
-C^A_i\, n_A = \frac{1}{g}\, \partial_i \varphi \left[P_{{\beta}_q}\frac{(1 +\cos \theta)}{2} + P_{{\beta}_p}\frac{(1-\cos \theta)}{2}\right]
-\frac{1}{g}\, L_{\alpha} \wedge \partial_i L_{\alpha}
\label{cfeten}
\end{equation}
\begin{equation}
P_{{\beta}_q} = \frac{1}{2}(\vec{\beta}_q+\vec{\beta}_p)\cdot \vec{T} + L_\alpha
\makebox[.5in]{,}
P_{{\beta}_p} = \frac{1}{2}(\vec{\beta}_q+\vec{\beta}_p)\cdot \vec{T} - L_\alpha \;.
\label{projectionb}
\end{equation}
The second term in eq. (\ref{cfeten}) generalizes the one obtained in the asymptotic region of an
$SU(2)$ 't Hooft-Polyakov monopole \cite{Manton}. This equation also puts in evidence the guiding centers of the junction, as well as the boundary conditions to be satisfied by ${\cal A}^A_i$. Besides the conditions ${\cal A}^A_i\to 0$ in the asymptotic region, and ${\cal A}^a_i\to 0$  in the asymptotic monopole region, these fields must compensate the singularities present at $\theta=0$ and $\theta=\pi$, in the first term of eq. (\ref{cfeten}), and the singularity at the origin, in the second. 

For the projections of the monopole current $H^q_{ij}$ in eq. (\ref{topo-curr}), we note that those with 
$\vec{\beta}\neq \vec{\beta}_q, \vec{\beta}_p$ vanish, as in that case $P_\beta$ is $x$-independent, while using eq. (\ref{projectionb}), 
\begin{equation}
\vec{\beta}_q|^{q'} H^{q'}_{ij}= -\frac{1}{g}\langle P_{\beta_q}, \partial_i L_{\alpha} \wedge \partial_j L_{\alpha} \rangle
\makebox[.5in]{,}
\vec{\beta}_p|^{q'} H^{q'}_{ij}= -\frac{1}{g}\langle P_{\beta_p}, \partial_i L_{\alpha} \wedge \partial_j L_{\alpha} \rangle
 \;.
\end{equation}
Then, taking into account eq. (\ref{elea}) we also obtain, $(\vec{\beta}_q+\vec{\beta}_p|^{q'} H^{q'}_{ij}=0$, so the
 nontrivial part of the monopole density current is contained in the projection along the root $\vec{\alpha}_{qp}$, 
\begin{equation}
\vec{\alpha}|^{q} H^{q}_{ij}/\alpha^2= -\frac{1}{g}\langle  L_\alpha, \partial_i L_{\alpha} \wedge \partial_j L_{\alpha} \rangle
\makebox[.5in]{,}
\vec{\alpha}=\vec{\alpha}_{qp}\;,
\end{equation} 
where we recognize the topological charge density for maps from $S^2 \to S^2$, namely, from the space of directions 
around a point into the space of local diagonal directions in a local $\mathfrak{so}(3)$ subalgebra.

\subsection{$N$-branch vortex-monopole junctions}

These junctions will confine a general color state of $N$ fundamental quarks, binding them to an appropriate monopole, forming a hybrid baryon. 
They can be defined by a mapping of the form,
\begin{equation}
S = e^{i(\varphi_1\, \vec{\beta}_1+\dots + \varphi_{N-1}\, \vec{\beta}_{N-1}) \cdot \vec{T}} W(x)\;,
\end{equation}
where the first factor is the one describing an $N$-branch normal junction.

On the disks $D_{{\rm c}_q}$ centered at ${\rm c}_q$, $q=1,\dots,N-1$, $S(x)$ behaves as,
\begin{equation}
S_{{\rm c}_q}(x) \sim  {\cal S}_{{\rm c}_q} \,  e^{i\chi_{{\rm c}_q}\, \vec{\beta}_{q} \cdot \vec{T}}\, W({\rm c}_q)\;.
\end{equation}
On the disk $D_{{\rm c}_N}$ centered at ${\rm c}_N$, when its radius $\epsilon$ gets smaller, we have,
\begin{equation}
S_{{\rm c}_N}(x) \sim  e^{i\varphi \, (\vec{\beta}_1 + \dots + \vec{\beta}_{N-1}) \cdot \vec{T}}\, W({\rm c}_N) = 
e^{i\chi_{{\rm c}_N}\, \vec{\beta}_N \cdot \vec{T}}\, W({\rm c}_N)\;.
\end{equation}
Then in order to have smooth mappings along the local Cartan directions, $n_q = S T_q S^ {-1}$ on $S^2$, we shall require
$W({\rm c}_1),\dots, W({\rm c}_N)$ to be either the identity or a Weyl reflection. The monopole charge is proportional to the sum $\frac{1}{2N} \sum_{{\rm c}} \vec{\beta}_{\rm c}$, which being the sum of $N$ weights of the fundamental representation, a reflected subset plus an unreflected one, is necessarily in the weight lattice of $Ad(SU(N))$.

\section{Center vortices and chains in pure YM}
\label{appa}

As we have seen, in the context of Yang-Mills-Higgs models, center vortices and junctions are interesting objects that could describe the states of the gluon field needed to confine quarks, and form the different normal and hybrid hadrons. To complete this picture, it is important having a description of heavy quark probes in the language of the proposed YMH model. This leads to another question, namely, why there could be some relationship between the YMH model and the pure QCD Yang-Mills theory it is supposed to describe.
Of course, this type of question corresponds to the open nonperturbative problem of confinement, so that an answer  from first principles is out of reach. What we could do is, based on some confinement scenarios in the lattice,  suggest some natural phenomenological choices, once the dual model is assumed to have the form given in eqs. (\ref{action}), (\ref{Higgs-po}). 

In the introduction, the description of the thin configurations in $SU(2)$ YM theory motivated the analisis of topological smooth configurations in the YMH model. Now, we can use the general properties of the latter configurations, discussed for a gauge grop $G$, 
to write those of the former, for a gauge group $G_{\rm e}$. Moreover, we would like to suggest, by means of a heuristic 
argument, a possible relationship among both sides.

Let us consider a YM theory with gauge group $G_{\rm e}$ and coupling constant $g_{\rm e}$, in $4D$ Euclidean spacetime, 
\begin{equation}
S^{\rm e}_{\rm YM}= \int d^4 x\, \frac{\varepsilon}{4} \langle F^{\rm e}_{\mu \nu}, F^{\rm e}_{\mu \nu}\rangle \;,
\label{YM-pure}
\end{equation}
for a gauge field $A^{\rm e}_\mu$ representing gluons. 
As is well-known, its analysis in the lattice supports scenarios dealing with center vortices, monopoles, and chains formed by them. For example, some observed properties of the quark potential, such as asymptotic linearity and $N$-ality\footnote{the fact that in $SU(N)$ YM theories the asymptotic string tension depends on the behavior of the quark representation under $Z(N)$.}, are well described by the ensemble of center projected configurations (see refs. \cite{greensite,book-G}). They are obtained by initially choosing a gauge such that the link variables $U_\mu(x)$ are as close as possible to some $Z(N)$ lattice configuration, then decomposing,
\begin{equation}
 U_\mu(x)=  {\cal P}_\mu(x)\, Z_\mu(x) \;,
\label{Ldec}
\end{equation}
where $Z_\mu(x) I$ is the center element of SU(N) closest to $U_\mu(x)$, and finally discarding the ``perturbative part''
${\cal P}_\mu(x)$,
\[U_\mu(x) \to Z_\mu(x)\;.\]
The obtained configurations essentially correspond to center vortex defects, located at plaquettes where $\prod Z_\mu(x)\neq 1$.
At low temperatures, center vortices percolate, and the average of the Wilson loop in the ensemble displays an area law, that is, confinement.
While this scenario is very good at describing $N$-ality, the stringlike behavior of the confining potential, that is, a nontrivial L\"uscher term associated with transverse fluctuations of the string, has not been observed in the center-projected data. 

In the lattice, monopoles and antimonopoles are also detected. Most of them have been found to be correlated with center vortex 
worldsheets to form chains, configurations that are considered very promising candidates to explain the different properties of the confining quark potential. Scenarios only involving an ensemble of monopoles have also proven successful in a number of respects.
In this case, the confining phase corresponds to a monopole condensate. For a detailed discussion of these topics, see \cite{greensite,book-G} and references therein.

Similar considerations to those given in \S \ref{topo-cent} and \S \ref{chains-comp} for the YMH model, here will lead to center vortices concentrated on worldsheets, that can concatenate looplike monopoles to form vortex-monopole thin junctions, or chains, in the language used in the lattice literature. Besides the dimensionality, there are important differences
with respect to the YMH configurations. At the classical level, the YM theory has no scales, at the quantum level, smeared  topological defects will be {\it postulated} to be relevant field configurations to compute averages. In this regard, a configuration of the form (\ref{ansatz}), with ${\cal A}^A_\mu \to {\cal P}^A_\mu$, $n_A \to u_A$, representing a perturbative and nonperturbative sector, is the analogous of the lattice decomposition (\ref{Ldec}), generalized to include all types of defect,
\begin{equation}
A^{\rm e}_\mu = ({\cal P}^{ A}_\mu -Z^{A}_\mu)\, u_A 
\makebox[.3in]{,}
Z^A_\mu = -(1/g_{\rm e})\,f^{\rm e}_{ABC}\, \langle u_B,\partial_i u_C \rangle 
\makebox[.3in]{,}
u_A = S T^{\rm e}_A S^{-1} \;,
\label{ansatz-e}
\end{equation}
where $S \in G_{\rm e}$ and the $T^{\rm e}_A$'s are generators of the Lie algebra $\mathfrak{g}_{\rm e}$. 
The field strength tensor is (cf. eq. (\ref{field-st})),
\begin{equation}
F^{\rm e}_{\mu \nu} = G^{{\rm e}\,A}_{\mu \nu}\, u_A
\makebox[.5in]{,}
G^{{\rm e}\,A}_{\mu \nu}=(F^A_{\mu \nu}({\cal P}) - F^A_{\mu \nu}(Z))
\;.
\label{field-st-e}
\end{equation}
Here, it is understood that quantities depending on the fields ${\cal P}_\mu$ and $Z_\mu$, involve the chromoelectric structure constants $f^{\rm e}_{ABC}$ and coupling $g_{\rm e}$. 

For example, a closed center vortex worldsheet $\Sigma$ is described by,
\begin{equation}
S = e^{i\chi\, \vec{\eta}\cdot \vec{T}^{\rm e}}
\makebox[.5in]{,}
\frac{1}{2\bar{N}}\, \vec{\eta} \in \Lambda(\tilde{G_{\rm e}^{\sf v}})\;,
\label{defcv}
\end{equation}
where $\chi$ is a multivalued function, changing by $2\pi$ when we go around a path linking $ \Sigma$.
Using eq. (\ref{F-cv}), we have,
\begin{equation}
-F^q_{\mu \nu}(Z) = \frac{1}{g_{\rm e}}\, \vec{\eta}|^q \, [\partial_\mu,\partial_\nu] \chi \;,
\end{equation}
and the vortex guiding centers are represented by a contribution to the dual tensor ${\cal F}^q_{\mu \nu}(Z) =\frac{1}{2}\, \epsilon_{\mu \nu \rho \sigma} F^q_{\rho \sigma}(Z)$,
\begin{eqnarray}
-{\cal F}^q_{\mu \nu}(Z) &=& \frac{2\pi}{g_{\rm e}}\, \vec{\eta}|^q \, \oint d^2 \sigma_{\mu \nu}\, \delta^{(4)} (x-\bar{y}(\sigma_1,\sigma_2))
\label{opencv}\;,
\end{eqnarray}
\begin{equation}
d^2 \sigma_{\mu \nu} = d\sigma_1 d\sigma_2\, \left(\frac{\partial \bar{y}_{\mu}}{\partial \sigma_1} \frac{\partial \bar{y}_{\nu}}{\partial \sigma_2} -\frac{\partial \bar{y}_{\mu}}{\partial \sigma_2} \frac{\partial \bar{y}_{\nu}}{\partial \sigma_1}\right)\;,
\end{equation}
where $d^2 \sigma_{\mu \nu}$ integrates over $\Sigma$, parametrized by $\bar{y}(\sigma_1,\sigma_2)$ (see \cite{engelhardt1}-\cite{lucho}).

For vortex-monopole junctions, the mapping $S$ has the general properties discussed in \S \ref{chains-comp}, so that, in the dual tensor, there is a sum over open center vortices attached to looplike monopole worldlines.
To identify the monopole guiding centers, note that for
each open vortex, we have,
\begin{eqnarray}
-\partial_\nu {\cal F}^q_{\mu \nu}(Z)|_{\rm vort}&=& \frac{2\pi}{g_{\rm e}}\, \vec{\eta}|^q \, \int d^2 \sigma_{\mu \nu}\, \partial_\nu \delta^{(4)} (x-\bar{y}(\sigma_1,\sigma_2))\nonumber \\
&=& \frac{2\pi}{g_{\rm e}}\, \vec{\eta}|^q \left( \oint_{C^+}  dy_\mu\, \delta^{(4)}(x-y) - \oint_{C^-}  dy_\mu\, \delta^{(4)}(x-y)  \right)\;, \nonumber
\end{eqnarray}
where $C^+$ ($C^-$) is the loop where the monopole (antimonopole) is localized. On the other hand, at say the monopole, the weights of the attached vortices must satisfy a condition analogous to eq. (\ref{possible}). That is,
for each monopole (antimonopole), the guiding centers are represented by a nonzero contribution to the divergence of the dual tensor,
\begin{equation}
-\, \partial_\nu {\cal F}^q_{\mu \nu}(Z)|_{\rm mon}= \pm \frac{2\pi}{g_{\rm e}}\, (\vec{\eta}_{{\rm c}_1}+\vec{\eta}_{{\rm c}_2}+\dots)|^q  \oint_{C}  dy_\mu\, \delta^{(4)}(x-y)\;,
\label{div-mon}
\end{equation}
\begin{equation}
\frac{1}{2\bar{N}}\, (\vec{\eta}_{{\rm c}_1}+\vec{\eta}_{{\rm c}_2}+\dots) \in \Lambda(Ad(G_{\rm e}^{\sf v}))\;.
\label{possible-e}
\end{equation}

In YM theory, when the vortex guiding centers leaving a monopole are superimposed, they form an unobservable open Dirac worldsheet. In this case,  the only objects expected in the ensemble are the monopole worldlines.

\subsection{Mesonic Wilson loop}

For a given irreducible representation of $G_{\rm e}$, the Wilson loop for external charges is,
\begin{eqnarray}
W_{{\cal C}}[A^{\rm e}]&=&(1/d_r)\, tr\, P \exp \left(ig_{\rm e}\oint_{{\cal C}} dx_\mu\, A^{\rm e}_\mu\right)\nonumber \\
&=&(1/d_r)\, tr\, [ S_f\, P \exp \left(ig_{\rm e}\oint_{{\cal C}} dx_\mu\, {\cal P}_\mu\right)S_i^{-1} ] \nonumber \\
&=& \mathfrak{z}({\cal C})\, W_{{\cal C}}[{\cal P}] \;,
\label{wloo}\end{eqnarray}
where $S_i$ and $S_f$ are the initial and final values of $S$ when we go around ${\cal C}$, a loop 
that represents a process where heavy quark antiquark probes are created, propagated and then annhilated. 
Here, we used that the product $ S_i^{-1} S_f$ must be an element $\mathfrak{z}({\cal C}) I$ in the center of $G_{\rm e}$, depending on ${\cal C}$, 
the distribution of defects, and the irreducible representation used for the external quarks.
Regarding this factor, let us initially consider it  in the presence of a closed  vortex  defined by eq. (\ref{defcv}). When the center vortex is linked, $\mathfrak{z}(\cal C)$ gives $e^ {i2\pi\, \vec{\eta} \cdot \vec{T}_{\rm e}} \in Z(G_{\rm e})$. This center element is a  number times the identity matrix, so it can be computed as the eigenvalue obtained when operating on any weight vector of the quark representation, labeled by a weight $\vec{w}_{\rm e}$. That is,
\begin{equation}
\mathfrak{z}({\cal C}) =  e^{i2\pi\, \vec{\eta} \cdot \vec{w}_{\rm e}\, L({\cal C}, \Sigma)} \;,
\label{IandII}
\end{equation}
where $L({\cal C},  \Sigma)$ is the linking number between ${\cal C}$ and $\Sigma$.
In addition, the linking number can be equated to the intersection number $I(S({\cal C}), \Sigma)$ between $S({\cal C})$, a surface whose border is ${\cal C}$, and the surface $ \Sigma$,
\begin{equation}
I(S({\cal C}), \Sigma)= \frac{1}{2} \int d^2 \tilde{\sigma}_{\mu \nu} \int_{\partial \Sigma} d^2 \sigma_{\mu \nu}\,
\delta^{(4)} (\bar{w}(s,\tau)-\bar{y}(\sigma_1,\sigma_2))\;,
\end{equation}
\begin{equation}
d^2 \tilde{\sigma}_{\mu \nu} = \frac{1}{2} \epsilon_{\mu \nu \alpha \beta}\,  d\tau\,  ds\, \left(\frac{\partial \bar{w}_{\alpha}}{\partial \tau} \frac{\partial \bar{w}_{\beta}}{\partial s} -\frac{\partial \bar{w}_{\alpha}}{\partial s} \frac{\partial \bar{w}_{\beta}}{\partial \tau}\right)\;,
\end{equation}
where $\bar{w}(s,\tau)$ is a parametrization of $S({\cal C})$ \cite{engelhardt1,reinhardt}. Then, introducing the source,
\begin{equation}
s_{\mu \nu}(x) = \int_{S({\cal C})} d^2 \tilde{\sigma}_{\mu \nu} \, \delta^{(4)}(x-\bar{w}(s,\tau))\;,
\end{equation}
and using eq. (\ref{opencv}), we obtain,
\begin{equation}
\mathfrak{z}({\cal C}) =  e^{- \frac{i}{2} \int d^4x\;  g_{\rm e} s_{\mu \nu}\vec{w}_{\rm e}\cdot  \vec{{\cal F}}_{\mu \nu}(Z)   } \;.
\label{inter-n}
\end{equation}

This representation can also be used for vortex-monopole junctions. In this case, there is also a (richer) concept of linking, as they are formed by monopoles and antimonopoles concatenated by center vortices to form closed objects. In addition, the intersection number can be applied to open surfaces, i.e.,
$S({\cal C})$ and the open vortices in the complex. Then, the center element $\mathfrak{z}({\cal C})$ can also be computed using eq. (\ref{inter-n}), that because of the flux matching between monopoles and center vortices, will give the same result for any possible $S({\cal C})$, as long as the monopoles do not touch $S({\cal C})$.

\subsection{$SU(N)$ baryonic Wilson loop}

Here, we shall consider $G_{\rm e}=SU(N)$, and the variable for a state formed by $N$ fundamental quarks. The 
baryonic potential can be measured from the average of the gauge invariant baryonic Wilson loop \cite{suganuma1}-\cite{cornwall}, \cite{cornwall1},
\begin{equation}
W_{Nq}[A^{\rm e}] = \frac{1}{N!}\;, \epsilon_{a_1 \dots a_N} \epsilon_{b_1 \dots b_N}\, U_{a_1b_1}(A^{\rm e},\Gamma_1) \dots U_{a_Nb_N}(A^{\rm e},\Gamma_N) \;,
\label{Nq}
\end{equation}
in a similar manner to the $q\bar{q}$ potential, that can be measured from the large distance behavior of the average 
$\langle W_{{\cal C}} \rangle$. The U-factors in eq. (\ref{Nq}) are the matrix elements of the holonomy,
\begin{equation}
U(A^{\rm e},\Gamma) = P \exp \left(ig_{\rm e}\int_{\Gamma} dx_\mu\, A^{\rm e}_\mu\right)\;,
\end{equation}
and the paths $\Gamma_1,\dots,\Gamma_N$  represent $N$ worldlines starting at a common point $x$, where the $N$ quarks are created, to a
common final point $y$, where they are annhilated. 

Using the composition property of the holonomy, we can compose the different quark worldlines with a common path $\Gamma^{-1}_0$, that
goes from $y$ to $x$, thus forming $N$ closed paths ${\cal C}_1=\Gamma_1\circ \Gamma^{-1}_0\dots {\cal C}_N=\Gamma_N\circ \Gamma^{-1}_0$, and write,
\begin{equation}
U(A^{\rm e},\Gamma)|_{ab}=U^{-1}(A^{\rm e}, \Gamma^{-1}_0)|_{ac} \, U(A^{\rm e},\Gamma \circ \Gamma^{-1}_0)|_{cb} \;.
\end{equation}
Now, for the different worldlines, the first factor corresponds to elements of the same matrix, so replacing in eq. (\ref{Nq}), and using 
$\det U^{-1} =1$, we get,
\begin{equation}
W_{Nq}[A^{\rm e}] = \frac{1}{N!}\, \epsilon_{a_1 \dots a_N} \epsilon_{b_1 \dots b_N}\, U_{a_1b_1}(A^{\rm e},{\cal C}_1) \dots U_{a_Nb_N}(A^{\rm e},{\cal C}_N) \;.
\label{Nq-c}
\end{equation}
In addition, for each closed path, we have,
\begin{equation}
S_{f}({\cal C})= \mathfrak{z}({\cal C})\, S_i\;,
\end{equation}
\begin{equation}
U (A^{\rm e},{\cal C}) = S_{f}({\cal C}) \,U ({\cal P},{\cal C})\, S^{-1}_{i} = \mathfrak{z}({\cal C})\, S_i\,U ({\cal P},{\cal C})\, S^{-1}_{i}\;.
\end{equation}
Then, replacing in eq. (\ref{Nq-c}), we obtain,
\begin{equation}
W_{Nq}[A^{\rm e}] = \mathfrak{z}({\cal C}_1)\dots \mathfrak{z}({\cal C}_N)\, W_{Nq}[{\cal P}]  \;,
\label{wdecs}
\end{equation}
where $\mathfrak{z}({\cal C})$ is given by eq. (\ref{inter-n}), using as $\vec{w}_{\rm e}$ a weight of the fundamental representation of $SU(N)$. This formula generalizes the one obtained in ref. \cite{cornwall1} for an ensemble of closed center vortices.

\subsection{Natural identifications}

The use of the decomposition (\ref{ansatz-e}), (\ref{field-st-e}) and the observables (\ref{wloo}), (\ref{wdecs}),
pose a difficulty,
when setting the fluctuations to zero, the thin defects have divergent action. On the other hand, the success of lattice scenarios involving defects could indicate that the proper ensemble is in fact formed by thick magnetic objects, with dimensional parameters depending on $\Lambda_{\rm QCD}$.
However, determining if this is the situation or not constitutes a difficult nonperturbative problem. At the perturbative level, the one-loop calculation of the effective action for a homogeneous magnetic diagonal background is lower than zero \cite{MS}, so magnetic configurations seem to be favored with respect to the classical vacuum. 
However, there is an imaginary part in the calculation \cite{sno}, implying a decay into off-diagonal gluons. This instability also  occurs for vortex backgrounds \cite{bordag,diakonov-center}, and it seems to be a general feature of magnetic bakgrounds.  
In ref. \cite{Lucho2}, we pointed out a possible way out for the instability problem, reexamining how to define the coupling between a center
vortex and quantum fluctuations. The idea is that fluctuations should not simply ``see'' the vortex as a diagonal background, but they should be coupled by the parametrization (\ref{ansatz-e}), where the presence of frame defects in the off-diagonal sector would 
change the counting of bound states of the fluctuation operator. 

The difficulties associated with a formulation from first principles of scenarios involving magnetic defects
in the continuum led to different models to compute averages, after postulating some possible properties of the ensembles. For example, an effective model of random surfaces, describing an ensemble of closed thick center vortex worldsheets, has been proposed in ref. \cite{engelhardt1}. These procedure implements  the lattice idea of center projection.  Arguments in favor of a Y-shaped potential between quarks were given in ref. \cite{cornwall}, assuming closed center vortices, and the statistical independence of the total linking numbers through ${\cal C}_1\dots {\cal C}_N$ in eq. (\ref{Nq-c}). In ref. \cite{Baker}, a dual Abelian YMH model for $SU(3)$ YM has been proposed. It contains a Higgs sector that provides an effective description of an ensemble of magnetic monopoles.   This type of model has been derived by assuming Abelian dominance, following a dualization procedure similar to that given in refs. \cite{polya,polyakov} for compact $QED(3)$ and $QED(4)$ (see also \cite{antonov}). 

A general ensemble should include the possibility of vortex-monopole junctions. However, it is no clear how to obtain an effective field theory when the defects are not one-dimensional. This is the case of junctions in
$3D$ Euclidean models \cite{Lucho3}. In $4D$, center vortices are two-dimensional worldsheets, 
for this reason, in order to suggest the natural choice of external sources and gauge group in the non Abelian YMH model, we shall 
only discuss the effect of monopoles, attached to unobservable Dirac worldsheets.

Let us sketch the main steps that can be followed after assuming Abelian dominance, and including mesonic and baryonic variables. Disregarding off-diagonal fluctuations in  ${\cal P}_\mu$, the holonomy  does not require path ordering. Then, the eigenvalues needed to compute traces and determinants involving $U({\cal P},{\cal C})$, can be simply obtained by acting on the weight vectors of the quark representation. In particular, the mesonic Wilson loop results, 
\begin{equation}
W_{{\cal C}}[A^{\rm e}]= (1/d_r) \sum_{w_{\rm e}} e^{\frac{i}{2} \int d^4x\;  g_{\rm e} s_{\mu \nu}\vec{w}_{\rm e}\cdot  
(\vec{{\cal F}}_{\mu \nu}({\cal P})-\vec{\cal F}_{\mu \nu}(Z))  } \;,
\end{equation}
where the sum is done over the weights of $\mathfrak{g}_{\rm e}$, corresponding to the irreducible representation considered. For a baryonic $SU(N)$ state formed by $N$ fundamental quarks, we get,
\begin{equation}
W_{Nq}[A^{\rm e}]=\frac{1}{N!}\, \sum_{P} e^{\frac{i}{2} \int d^4x\;  g_{\rm e} (s^1_{\mu \nu} \vec{w}_{p_1}+\dots + s^N_{\mu \nu}\vec{w}_{p_{N}})\cdot  (\vec{{\cal F}}_{\mu \nu}({\cal P})-\vec{\cal F}_{\mu \nu}(Z))  }  \;,
\end{equation}
where the sum is done over the $N!$ permutations of the weights of the fundamental representation. 

The different terms in these expressions are expected to have the same average, after integrating over
the ensemble with an appropriate phenomenological measure, and over the gauge fields, using the weight,
\begin{equation}
e^{-\int d^4x\, \frac{\varepsilon}{4}\vec{G}^{{\rm e}}_{\mu \nu}\cdot \vec{G}^{{\rm e}}_{\mu \nu}}=e^{-\int d^4x\, \frac{\varepsilon}{4}(\vec{F}_{\mu \nu}({\cal P}) - \vec{F}_{\mu \nu}(Z))^2}\;.
\end{equation}
Linearizing the kinetic term, both averages have the form,
\begin{eqnarray}
\langle W\rangle &=&
\left\langle e^{-\int d^4x\, \frac{\zeta}{4} (\vec{\Lambda}_{\mu \nu}-\vec{J}_{\mu \nu})^2} \,
e^{\frac{i}{2}\int d^4x\, (1/2\bar{N})\, \vec{\Lambda}_{\mu \nu} \cdot (\vec{{\cal F}}_{\mu \nu}({\cal P})-\vec{{\cal F}}_{\mu \nu}(Z))}\right\rangle 
\;,
\label{Wcyn}
\end{eqnarray}
$$
\vec{J}_{\mu \nu} = \left\{ \begin{array}{lll}
g_{\rm e} \vec{\beta}_{\rm e}s_{\mu \nu} \;, &{\rm for}& W=W_{{\cal C}} \\
g_{\rm e} \vec{\beta}_1 s^1_{\mu \nu}  + \dots +  g_{\rm e} \vec{\beta}_N\, s^N_{\mu \nu}  \;, &{\rm for} & W=W_{Nq}\;,
\label{ext-sour}
\end{array} \right.
$$
with  $\zeta =  1/4\bar{N}^2\varepsilon$. In eq. (\ref{Wcyn}), the integration variable $\vec{\Lambda}_{\mu \nu}$, which is coupled to
dual field strength tensors, has been rescaled so as to work with external currents having fluxes with normalization 
$g_{\rm e}\vec{\beta}=2\bar{N}\vec{w}$. 
To simplify the discussion, we shall initially assume that monopoles are singular pointlike objects. Note that the variable ${\cal P}_\mu$ is not required to satisfy any boundary condition on the unobservable Dirac worldsheets, so we can directly integrate to obtain the constraint,
\begin{equation}
\epsilon_{\mu \nu \rho \sigma} \partial_\nu \vec{\Lambda}_{\rho \sigma}=0 \;.
\end{equation}
Its solution leads to the dual representation,
\begin{eqnarray}
\langle W\rangle &=&
\left\langle e^{-\int d^4x\, \frac{\zeta}{4} (\vec{G}^{\,\rm P}_{\mu \nu}- \vec{J}_{\mu \nu})^2} \,
e^{-i\int d^4x\, \vec{ A}^{\,\rm P}_{\mu} \cdot \, \partial_\nu \vec{{\cal F}}_{\mu \nu}(Z)}\right\rangle 
\;,
\label{apro}
\end{eqnarray}
where $\vec{G}^{\,\rm P}_{\mu \nu}=\partial_\mu \vec{A}^{\,\rm P}_\nu -\partial_\nu \vec{ A}^{\,\rm P}_\mu$, and the index P refers to fields in the Abelian projected case. The ensemble average is not well defined unless some phenomenological dimensional parameters in the integration measure be introduced. Different choices for the string tension and interactions among the monopoles will lead to a variety of effective field models,
 exhibiting different phases \cite{bar-sam}-\cite{Kleinert-book}. 
Here, our main concern is only to suggest some natural identifications, which can be even seen after turning off the interactions.

Using eq. (\ref{div-mon}), including the sum over the different monopole worldlines, we see that the dual vector field 
$\vec{A}^{\,\rm P}_\mu$ is minimally coupled to particles whose charges are given by $(2\pi/g_{\rm e})$ times an element of the lattice $ \Lambda(Ad(G_{\rm e}^{\sf v}))$. Considering the elementary monopoles in eqs. (\ref{div-mon}), (\ref{possible-e}), which correspond to the 
roots $\vec{\alpha}_{\rm e}^{\,\sf v}$ of $\mathfrak{g}_{\rm e}^{\sf v}$, we have,
\begin{equation}
\langle W\rangle =
\int [DA_{\rm P}]\, e^{-\int d^4x\, \frac{\zeta}{4} (\vec{G}^{\,\rm P}_{\mu \nu}-\vec{J}_{\mu \nu}  )^2} \,
Z_m
\makebox[.5in]{,}
Z_m=Z_{{\alpha}_1} Z_{\alpha_2}\dots\;,
\end{equation}
\begin{equation}
Z_\alpha = \sum_N \frac{1}{N!} \prod_{k=1}^N\, [Dx_k]\, e^{-m \sum_{k=1}^N L_k}\, e^{i \frac{2\pi}{g_{\rm e}} \int_{C_k} dx_\mu\,  
\vec{\alpha}\cdot \vec{ A}^{\,\rm P}_{\mu} }
\label{zalpha}
\;,
\end{equation}

Here, we have set the interaction among the monopoles to zero, so that $Z_\alpha$, the partition function for  monopole loops with charge $\vec{\alpha}^{\, \sf v}_{\rm e}$, is obtained by summing over the number of loops, and for each sector, path integrating over all possible loop shapes with the measure $[Dx_k]$. To avoid overcounting, only the positive roots need to be considered.  
The configurations are weighted by the interaction in eq. (\ref{apro}), between the monopole loops and the projected dual field, as well as the cost to have a loop with length $L$, depending on a phenomenological parameter $m$.

Now, as shown in refs. \cite{HSi,bar-sam}, the right hand side in eq. (\ref{zalpha}) is a functional determinant for a vacuum to vacuum amplitud in ``particle'' representation. This means, that $Z_\alpha$ can be cast in the form,
\begin{equation}
Z_\alpha = \int [D\phi_\alpha] [D\bar{\phi}_\alpha]\, e^{-\int d^4x\, \bar{\phi}_\alpha [-D_\alpha^2 +m^2]\phi_\alpha}
\makebox[.3in]{,}
D^{\alpha}_\mu=\partial_\mu-i\frac{2\pi}{g_{\rm e}} \vec{\alpha}^{\,\sf v}_{\rm e}\cdot\vec{ A}^{\,\rm P}_{\mu}\;.
\end{equation}
where $\phi_\alpha$ is a complex field associated with a root $\vec{\alpha}^{\,\sf v}_{\rm e}$. In this respect, including  
density-density monopole interactions, some effective models have been obtained where a SSB phase is observed for $m^2<0$ \cite{antonov}.

Summarizing, when  considering Abelian projected configurations, and after turning off the interactions, the Wilson loop is,
\begin{equation}
\langle W\rangle =
\int [DA_{\rm P}]  [D\phi_{\alpha}] [D\bar{\phi}_\alpha]\    \, e^{-\int d^4x\, [\frac{\zeta}{4} (\vec{G}^{\,\rm P}_{\mu \nu}- \vec{J}_{\mu \nu} )^2+\sum_\alpha
\left[ | D^\alpha_\mu \phi_\alpha|^2 +m^2 \bar{\phi}_\alpha\phi_\alpha\right] } \;.
\label{Wgene}
\end{equation}
In the case of the mesonic Wilson loop, that has been written for a general $\mathfrak{g}_{\rm e}$, $\vec{\alpha}^{\,\sf v}_{\rm e}$ sums over the positive roots of $\mathfrak{g}_{\rm e}^{\sf v}$. For the $SU(N)$ baryonic Wilson loop, the sum is over the $N(N-1)/2$ positive roots of $\mathfrak{su}(N)$.

Then, for the class of models we presented in \S \ref{YMHm}, it is natural to introduce the  quark sources in the form, 
\[
S= \int d^{4} x\, \left(\frac{\zeta}{4} \langle F_{\mu \nu}-J_{\mu \nu}, F_{\mu \nu}-J_{\mu \nu}\rangle + \frac{1}{2} \langle D_\mu \psi_A , D_\mu \psi_A \rangle +
V_{\rm Higgs}(\psi_A)\right)\;,
\]
and require that this action, when restricted to Abelian configurations, coincide with that derived for the  Abelian projected effective model.
Considering,
\begin{equation}
A_\mu= \vec{A}^{\, \rm P}_\mu \cdot  \vec{T}
\makebox[.5in]{,}
J_{\mu \nu}  =\vec{J}_{\mu \nu}|^q\,T_q 
\end{equation}
\begin{equation}
\frac{\psi_\alpha + i \psi_{\bar{\alpha}}}{\sqrt{2}} =\phi_{\alpha}\, E_\alpha 
\makebox[.5in]{,} 
\frac{\psi_\alpha - i \psi_{\bar{\alpha}}}{\sqrt{2}} =\bar{\phi}_{\alpha} \,E_{-\alpha}\;,
\label{pf}
\end{equation}
the covariant derivatives in $S$ depend on the roots of $\mathfrak{g}$ and the coupling $g$, 
thanks to the fact the Higgs fields are in the adjoint, while
in eq. (\ref{Wgene}), they depend on the roots of $\mathfrak{g}^{\,\sf v}_{\rm e}$ and the coupling $2\pi/g_{\rm e}$.
In addition, the effective Abelian model in eq. (\ref{Wgene}) contains as many complex fields as positive roots of the Lie algebra. Then, from this perspective, it is natural identifying,
\begin{equation}
\mathfrak{g}=\mathfrak{g}_{\rm e}^{\sf v}
\makebox[.5in]{,}
g=\frac{2\pi}{g_{\rm e}}\;,
\label{ident}
\end{equation}
so when the interactions are turned off, the sector of off-diagonal flavors in the Abelian projected action $S$ reduces to the exponent in eq. (\ref{Wgene}), with $\mu^2 =m^2$.

\subsection{Induced topological configurations}

Note that the parameter $\zeta$ does not change the values of the center vortex and monopole charges of the YMH model discussed in \S \ref{topo-cent} and \S \ref{chains-comp}.
They are still given by eqs. (\ref{solution}), (\ref{equiv-vc}), and (\ref{cv-ch}), so after the identification (\ref{ident}), that implies
$\tilde{G}^{\sf v}=\tilde{G}_{\rm e}$, the vortex charges become electric charges,
\begin{equation}
\vec{Q}_v = g_{\rm e}\, \vec{\beta} \makebox[.5in]{,}
\frac{1}{2\bar{N}}\, \vec{\beta}  \in \Lambda(\tilde{G}_{\rm e})\;.
\end{equation}
These are precisely the fluxes that can match those of the external quark sources in eq. (\ref{ext-sour}). 

In particular, in a  process where a $q\bar{q}$ pair, or a $q\dots q$ baryon, is created at $\tau_i$ in the far past, the charges are static at intermediate times, and then they  are annhilated at $\tau_f$, in the far future, the parametrization of $S({\cal C})$ can be done as follows, 
\begin{equation}
\bar{w}(\tau,s)=(\tau,  \bar{x}(s))
\makebox[.5in]{,}
\tau_i < \tau < \tau_f
\makebox[.3in]{,}
\bar{x} \in R^3 \;,
\end{equation}
where $\bar{x}(s)$ joins the quark and the antiquark. This leads to static sources,
\begin{equation}
s_{0 i}(x)=0\makebox[.3in]{,} s_{ij}(x) = -\epsilon_{ijk}\, \int ds\,  \frac{d\bar{x}_k}{ds}\, \delta^{(3)}(x-\bar{x}(s))
\makebox[.5in]{,}
x \in R^3\;.
\end{equation}
Changing the shape of $\bar{x}(s)$ amounts to changing that of $S({\cal C})$, an operation that should have no observable effects. 
In other words, $\bar{x}(s)$ can be thought of as a Dirac string where only the endpoints (the quark locations) are observable.
If desired, for the mesonic Wilson loop we can use a pair of Dirac strings, 
one running from $-z_0$ to $-\infty$, the other running from $+\infty$ to $+z_0$. 
That is, $-\vec{K}_{ij}$ represents a flux $\vec{\beta}_{\rm e}$ leaving (entering) the quark (antiquark) position at $-z_0$ ($+z_0$).
For the baryonic Wilson loop, as $\vec{\beta}_1+\dots\vec{\beta}_N=0$, we can choose the $N$ Dirac strings  running from the quark positions to $\infty$, so that $-\vec{K}_{ij}$ represents fluxes $\vec{\beta}_1,\dots,\vec{\beta}_N$, leaving each one of the quarks.

The energy functional for (dual) magnetic configurations in the presence of static quarks is
\begin{equation}
E= \int d^{4} x\, \left(\frac{\zeta}{4} \langle F_{ij}-J_{ij}, F_{ij}-J_{ij}\rangle + \frac{1}{2} \langle D_i \psi_A , D_i \psi_A \rangle +
V_{\rm Higgs}(\psi_A)\right)\;,
\end{equation}
and its minimization leads to eq. (\ref{YMF}), with the substitution, 
\begin{equation}
 F_{ij}\to \zeta (F_{ij}-J_{ij})\;,
\end{equation}
while eq. (\ref{YMP}) is unaltered.
Now, we are interested in solutions that have smooth energy density, in spite of the singular term $J_{ij}$. 
Writing $A_\mu$ and $\psi_A$ in terms of ${\cal A}^A_i$, $n_A$, and $h_{AB}$, as in eqs. (\ref{ansatz}), (\ref{hyS}),  and taking into account eq. (\ref{field-st}), we then require a regular,
\begin{equation}
F_{ij}-J_{ij} =  F^A_{i j}({\cal A}) \, n_A-F^A_{ij}(C)\,  n_A- J_{ij}\;.
\label{lt}
\end{equation}

In the mesonic case, 
this requirement can be fulfilled by inducing a finite length vortex with guiding centers $\bar{x}_{\rm c}(s)$, running from the quark to the antiquark, and weight $\vec{\beta}$ equal to the quark's weight $\vec{\beta}_{\rm e}$. That is, by inducing a configuration parametrized by, 
\begin{equation}
S=e^{i\chi\, \vec{\beta}_{\rm e}\cdot \vec{M}}
\makebox[.5in]{,}
R= e^{i\chi\, \vec{\beta}_{\rm e}\cdot \vec{M}}\;,
\label{abescreen}
\end{equation}
where $\chi$ is a multivalued function that changes by $2\pi$, when the closed curve formed by the composition of 
$\bar{x}_{\rm c}(s)$ and the string $\bar{x}^{-1}(s)$ is linked.
Using the relation between couplings in eq. (\ref{ident}), the last two terms in eq. (\ref{lt}) become,
\begin{eqnarray}
(1/g) [\partial_i,\partial_j]\chi\, \vec{\beta}_{\rm e}\cdot \vec{M}+ g_{\rm e}\epsilon_{ijk}\, \int ds\,  \frac{d\bar{x}^k}{ds}\, \delta^{(3)}(x-\bar{x}(s))\, \vec{\beta}_{\rm e}  \cdot \vec{M}=&&\nonumber \\
=g_{\rm e}\epsilon_{ijk}\, \int ds\,  \frac{d\bar{x}^k_{\rm c}}{ds}\, \delta^{(3)}(x-\bar{x}_{\rm c}(s))\, \vec{\beta}_{\rm e}  \cdot \vec{M}\;. &&
\end{eqnarray}
This sum vanishes at the Dirac string $\bar{x}(s)$, so ${\cal A}_\mu$ must be smooth there and, as a consequence, no special condition for the profile functions $h_{AB}$ is required at $\bar{x}(s)$.
On the other hand, at the vortex guiding centers, 
$\vec{J}_{ij}$  vanishes. Then, when $x\to \bar{x}_{\rm c}(s)$, the diagonal components of ${\cal A}_\mu$ along $\vec{\beta}_{\rm e}$ must tend to $-\frac{1}{g}\partial_i \varphi\, \vec{\beta}_{\rm e} \cdot \vec{T}$, for the sum of the first two terms in eq. (\ref{lt}) be smooth, In addition, when $x\to \bar{x}_{\rm c}(s)$, the Higgs profiles coupled to these components must tend to zero. These boundary conditions must be interpolated with those at $\infty$, so the energy density will be distributed around $\bar{x}_{\rm c}(s)$.

For the baryonic Wilson loop, we have seen that the external sources are associated with the $N$ different weights $\vec{\beta}_1, \dots, \vec{\beta}_N$ of the fundamental representation of $SU(N)$. In this case, a regular energy density can be obtained by inducing a configuration with Y-shaped finite guiding centers that, combined with the external sources, are described by the map  $S$ given in eq. (\ref{normalY}). Again, because of the external sources, the usual regularity conditions only need to be imposed at the guiding centers, thus localizing the energy density on a finite Y-junction.

Then, when introducing external quarks, the equations of motion are those given in (\ref{YMF}), (\ref{YMP}) but with modified boundary conditions. A similar situation applies to the $SU(N)$ hybrid meson. It can be induced by only using the regularity conditions, discussed
in \S \ref{two-branch}, inside a sphere $S^2$ around the origin, and using a regular ${\cal A}^A_\mu$ outside. The apparent singularities in the energy density outside $S^2$ will cancel against appropriate external sources, corresponding to a quark and antiquark with charges $\vec{\beta}_q$, -$\vec{\beta}_p$ placed at $-z_0$ and $+z_0$, respectively. 

The search for the minimum energy must also consider a minimization with respect to the guiding centers. For example, for fundamental quarks in $SU(3)$, when considering a $q\bar{q}$ pair or  three quarks on the corners of an equilateral triangle, this step is expected to be unnecessary, if the guiding centers form a straight segment joining the quarks, and the Y-junction branches are oriented at angles of $120^{\circ}$, respectively. 

Summarizing, the topological features of the different configurations, the boundary conditions needed to have a regular energy density, as well as the dimensional scales associated with the diagonal and off-diagonal Higgs sectors, are expected to lead to finite energy solutions
to the equations of motion. However, it is important to emphasize that care must be taken so as to avoid spurious solutions.  For example, in $SU(N)$,  following the steps 
we did for a $q\bar{q}$ pair, using a weight $\vec{\beta}_{\rm e}$ in the adjoint, it seems that a string confining adjoint quarks could be formed. The point is that we should not insist on screening the external sources only using the Abelian phase in eq. (\ref{abescreen}).  When imposing the boundary conditions, all types of  $R$ mappings should be considered. A similar situation was discussed at the end of \S \ref{wc}. In fact, when the sources are in the adjoint, a non Abelian phase that screens the external sources, and is smooth on the region between the charges, will be induced, thus implying a vanishing string tension for adjoint quarks, in accordance with $N$-ality.

\section{Conclusions}
\label{conc}

In this article, we analyzed a class of Yang-Mills models containing adjoint Higgs fields, with $SU(N)$ symmetry spontaneously broken down to $Z(N)$, and showed they  contain a rich variety of topological objects. These include center vortices, $N$-branch junctions exclusively formed by them, and junctions where different center vortices are interpolated by  monopole-like configurations. In the context of dual superconductor models, the first one  represents the string that confines a color neutral quark/antiquark pair,  to form a meson.
The second possibility represents  the generalized Y-shaped potential that confines  a color neutral $N$-quark state, to form a baryon. The third,  that contains correlated monopoles, is a new kind of configuration that naturally arises
in these models. The simplest
can be thought of as a gluon field state containing a colored valence gluon represented by the monopole.
They could confine colored quark states to form an overall color neutral hybrid object. For example, a junction formed by a pair of center vortices with different charges, interpolated by a monopole, can confine a nonsinglet quark/antiquark pair, 
to form a hybrid color neutral $qg\bar{q}'$ meson. Similarly, hybrid baryon states can be formed by $N$-branch vortex-monopole junctions.

These states are indeed allowed by QCD, existing some candidates in the observed hadron spectrum, lattice evidence in favor of them, as well as experimental collaborations searching for this type of  gluonic excitation.
  
Although the main interest is on $\mathfrak{su}(3)$, we have found it convenient to settle the discussion in the  framework of a connected compact simple Lie algebra $\mathfrak{g}$, and use its 
general structure. In our model, vacuum configurations in the SSB phase are labeled by $n_A$, a local basis for $\mathfrak{g}$, that admits a Cartan decomposition into  diagonal ($n_q$) and off-diagonal ($n_a$) local sectors. Equivalently, they can be labeled by a mapping $R\in Ad(G)$.
 
As is well known, when the center of the universal covering group $\tilde{G}$ is nontrivial, stringlike center vortices  can be formed, as the first homotopy group of the space of vacua is given by $\Pi_1(Ad(G))=Z(\tilde{G})$. On  the other hand, {\it isolated} monopole-like objects cannot, as $\Pi_2(Ad(G))=0$. Here, we have emphazised the fact that mappings from $S^2$, the sphere around a point, into the space of vacua, only measures the possibility of having isolated objects. That is, the trivial $\Pi_2$ of a compact group does not forbid monopole-like objects attached to center vortices to form junctions. In such a configuration, on the sphere around the monopole, there would be regions of false vacuum, namely, the center vortex cores. As we have shown, these junctions can be formed indeed, with the monopole contained in the local diagonal sector of the Lie algebra. They are described by the exact homotopy sequence,
$$
0=\Pi_2(Ad(G))\rightarrow \Pi_2 \left(D\right)\rightarrow \Pi_1(Ad\, (H)) \rightarrow \Pi_1(Ad\, (G))=Z(\tilde{G})\;,$$ where $D=Ad(G)/Ad(H)$ , and $H$ is the subgroup of $G$ leaving the diagonal generators $T_q$ invariant. In this sequence, the manifold $D$ can be identified with the local diagonal sector $n_q=ST_qS^{-1}$, $S\in G$. 
The manifold $H$ is associated with the center vortices, as when we get closer to their guiding centers they are essentially contained in a local Abelian subgroup. 
The $S^1$ in $\Pi_1(H)$ refers to the composition of loops around the different center vortex cores crossing the surface $S^2$, where $\Pi_2(D)$ is being analyzed.
Then, when this composition is required to be trivial in $Ad(G)$, an identification between $\Pi_2(D)$ and $\Pi_1(H)$ is established. This means that configurations exist where a monopole along the diagonal sector is attached to center vortices, as long as the charges of the vortices, which 
are essentially weights of  the universal covering of the dual group $G^{\sf v}$, add up to a weight of $Ad(G^{\sf v})$. In turn, the weights of  $Ad(G^{\sf v})$ become the possible monopole charges, and the absence of isolated monopoles can be interpreted as the confinement of valence gluons.

In the case of $SU(N)$, the simplest center vortices are known to be labeled by weights of the fundamental representation. On the other hand, the simplest monopoles are labeled by the roots, which  can be written as the difference of a pair of fundamental weights, those associated with the pair of interpolated center vortices. 

As a possible model supporting the different topological objects, we proposed one containing as many adjoint Higgs fields $\psi_A$ as the dimension of the Lie algebra, and constructed a potential displaying flavor symmetry with global group $Ad(G)$. Its terms were obtained by considering Higgs field products in the Lie algebra and forming invariants up to quartic order in the fields. When the cubic term is nonzero, as the square mass term is lowered, a first order transition to a SSB phase is obtained. When the parameter of the cubic term is nullified, there is a second order line.

In the SSB phase, the minimization of the potential turns the set of fields $\psi_A$ into a local basis $n_A$, so that the previous analysis concerning the possible topological objects  follows. The mapping $R\in Ad(G)$, associated with $n_A$, is a non Abelian phase whose defects determine the guiding centers of the smooth topological objects, as well as the boundary conditions to be imposed on the gauge and Higgs fields. 

For $SU(2)$ and $SU(3)$ center vortices, we explicitly showed how to close the field equations for a reduced set of profile functions. The equation for  $a(\rho)$, the gauge field profile along the center vortex weight, is the same encountered for the Nielsen-Olesen vortex. Here, the Higgs profile $h(\rho)$ is the one associated with those off-diagonal components that are rotated by diagonal transformations, with axis along the vortex weight, so it must vanish at the origin. The field equation for $h(\rho)$ is a modified Nielsen-Olesen equation,
containing couplings to profile functions for the Higgs fields that are not rotated. 

For general $SU(N)$, we wrote the non Abelian phase describing the guiding centers of a junction formed by a pair of fundamental center vortices, with weights $\vec{w}_q$ and $\vec{w}_p$, attached to a monopole. In this case, we showed how the monopole is contained in $L_\alpha= \vec{\alpha}|^q n_q/\alpha^2 $, $\vec{\alpha}=\vec{w}_q-\vec{w}_p$, the generator of a local 
$\mathfrak{so}(3)$ algebra. For the gauge field, we put in evidence the presence of a center vortex in the positive (resp. negative) $z$-axis with weight $\vec{w}_q$ (resp. $\vec{w}_p$), and the monopole term $L_{\alpha} \wedge \partial_i L_{\alpha}$. We also showed that the monopole charge is the topological charge for the mapping $S^2 \to \{L_{\alpha}\}$.

The diagonal Higgs fields are important to set a scale for the monopoles, giving a smooth transition from the asymptotic monopole region, where the diagonal sector does not contribute, to the center of the monopole, where the field profiles accompanying 
the singular $L_\alpha$ must tend to zero.

The general picture is completed with a heuristic discussion about what could be the relation between the YMH model with gauge group $G$ and pure YM with gauge group $G_{\rm e}$.  The natural choice of external sources in the YMH model, to describe quarks in different irreducible representations of $G_{\rm e}$, is also addressed. 
The idea is that configurations involving monopoles, center vortices and vortex-monopole junctions (chains, in lattice language) also appear in pure YM theories defined on the lattice. The Abelian projection, leading to ensembles of monopoles, and the center projection, where only the center vortex degrees are maintained in the averages, proved to be very successful to describe some of the properties of the confining force. Chains are also candidates to explain the different aspects of this force. 

In continuum Euclidean YM theory, the different projections can be simulated by considering ensembles of magnetic defects, characterized by a phenomenological measure that contains dimensional parameters. In $4D$, the monopole component is formed by looplike defects, placed at the border of observable center vortex worldsheets, and unobservable Dirac worldsheets. As is well known, in Abelian projection scenarios, an effective description of this component is expected to be a {\it field} theory. This comes about as the monopole ensemble contains a sum over different numbers of closed worldlines, that is, a second quantized field theory. As the possible elementary charges of YM monopoles are roots spanning the lattice $Ad(G^{\sf v}_{\rm e})$, the Abelian projected effective model contains as many charged fields as the number of positive roots. These modes were identified with the off-diagonal flavors of the Abelian projected YMH model, when defined with adjoint Higgs fields transforming in $Ad(G^{\sf v}_{\rm e})$, that is, it is natural identifying $G=G^{\sf v}_{\rm e}$.

In this regard, it would be interesting to look for the ensemble whose large distance behavior is described by the complete YMH model we proposed, although obtaining an effective model outside the Abelian projection assumption is a difficult task, as well as dealing with ensembles of vortex-monopole junctions in $4D$.

Because of the identification $G=G^{\sf v}_{\rm e}$, the YMH center vortex charges become weights of  the universal covering of $G^{\sf v}=G_{\rm e}$.  In addition, in dual language, we have shown that the mesonic and baryonic YM Wilson loops  become naturally represented by external pointlike monopole sources, with weights of the irreducible representation of $G_{\rm e}$ where the quark probes are defined. That is, the quark sources can be matched and confined by center vortices, junctions formed by them, and junctions containing interpolating monopoles. In the YMH model, these configurations are smooth, in contrast to what happens in the YM theory. They are solutions to the classical equations of motion, representing normal and hybrid mesonic and baryonic states. Further research on this direction, performing a numerical analysis of the equations, will permit to obtain the model's potential between a quark/antiquark pair, for various irreducible representations, and the field distribution of confining strings and valence gluons inside hadrons.

\section*{Acknowledgements}

I am grateful to Nick Manton for very helpful discussions, and David Tong, for raising interesting questions that motivated this
work. I would also like to thank for their hospitality the DAMTP, during the completion of this project, and the ICTP, during a visit. The Conselho Nacional de Desenvolvimento Cient\'{\i}fico e Tecnol\'{o}gico (CNPq-Brazil), the Funda\c c\~ao de Amparo a Pesquisa do Estado do Rio de Janeiro (FAPERJ), and the Proppi-UFF are acknowledged for the financial support.

%%%%%%%%%%%%%%%%%%%%%%%%%%%%%%%%%%%%%%%%%%%%%%%%%%%%%%%%%%%%%%%%%%%%%%
%%%%%%%%%%%%%%%%%%%%%%%%%%%%%%%%%%%%%%%%%%%%%%%%%%%%%%%%%%%%%%%%%%%%%%
%%%%%%%%%%%%%%%%%%%%%%%%%%%  References  %%%%%%%%%%%%%%%%%%%%%%%%%%%%%
%%%%%%%%%%%%%%%%%%%%%%%%%%%%%%%%%%%%%%%%%%%%%%%%%%%%%%%%%%%%%%%%%%%%%%
%%%%%%%%%%%%%%%%%%%%%%%%%%%%%%%%%%%%%%%%%%%%%%%%%%%%%%%%%%%%%%%%%%%%%%

\bibliographystyle{h-physrev4}

\end{document}